\documentclass[12pt,a4paper]{article}
\usepackage{aip1}
\usepackage{amssymb}
\usepackage{graphicx}

\newtheorem{theorem}{Theorem}[section]
\newtheorem{prop}[theorem]{Proposition}
\newtheorem{lemma}[theorem]{Lemma}

\newcommand{\proof}{\noindent {\bf {Proof}} \smallskip \newline }

\newcommand{\be}{\begin{equation}}
\newcommand{\ee}{\end{equation}}

\newcommand{\bea}{\begin{eqnarray}}
\newcommand{\eea}{\end{eqnarray}}

\begin{document}

 
 
\begin{centering}
\vspace{.05in}
{\Large {\bf Existence theorems for hairy black holes in 
${\mathfrak {su}}(N)$ Einstein-Yang-Mills theories.}}
\vspace{.2in}

N.E. Mavromatos$^{*}$ 
and E. Winstanley \\
\vspace{.05in}
University of Oxford, Theoretical Physics, 
1 Keble Road OX1 3NP, U.K. \\

\vspace{.2in}
{\bf Abstract} \\
\vspace{.05in}
\end{centering}
We establish the existence of hairy black holes 
in ${\mathfrak {su}}(N)$
Einstein-Yang-Mills theories, described by $N-1$  parameters,
corresponding to the nodes of the gauge field functions.

\vspace{.1in}
\begin{centering}
PACS numbers: 04.40.Nr, 04.70
\end{centering}

\vspace{2.5in}

\begin{flushleft} 
December 1997 \\
\end{flushleft} 


\newpage

\section{Introduction}
\label{intro}
Interest in non-Abelian Einstein-Yang-Mills theories was sparked by 
the discovery of particle-like \cite{bartnik} and 
non-Reissner-Nordstr\"om (``hairy'') black hole \cite{bizon}
solutions when the gauge group is ${\mathfrak {su}}(2)$.
Since then a plethora of hairy black holes, possessing non-trivial
geometry and field structure outside the event horizon, have been
found (see, for example, \cite{hairy}), including 
coloured black holes in ${\mathfrak {su}}(N)$ 
Einstein-Yang-Mills theories \cite{klei1,klei2}.
Many of these objects have as a fundamental requirement for the
existence of hair, a non-Abelian gauge field, often coupled to 
other fields, such as a Higgs field \cite{EM}.
As in \cite{EM}, the existence of gauge hair is not surprising in
itself, since the gauge field force is long-range.
However, the non-Abelian nature of the field is important 
for evading the no-hair theorem, for example, for scalar fields
coupled to the gauge field.
The vast majority of these solutions have been found only numerically,
with analytic work in this area at present being limited to 
an extensive study of the ${\mathfrak {su}}(2)$ case
\cite{bfm,smoller,sw}, and some analysis of the field equations
for general $N$ \cite{kunzle,kunz1}.

\bigskip
In this paper we continue this analysis of the coupled 
Einstein-Yang-Mills equations for an ${\mathfrak {su}}(N)$
gauge field, and prove analytically the existence of 
``genuine'' hairy black hole solutions for every $N$.
By a ``genuine'' ${\mathfrak {su}}(N)$ black hole, we mean a 
solution which is not simply the result of embedding a 
smaller gauge group in ${\mathfrak {su}}(N)$.
As in the ${\mathfrak {su}}(2)$ case, the solutions are labeled
by the number of nodes of each of the $N-1$ non-zero functions
required to describe the gauge field.
We shall prove that for each integer $n_{1}$, there are an infinite
number of sequences of integers 
$n_{N-1}\ge n_{N-2}\ge \ldots \ge n_{1}$
corresponding to black hole solutions.
It is to be expected that in fact {\em {every}} such sequence
corresponds to a black hole solution, but unfortunately we are unable
to prove this analytically, although we shall present a
numerically-based argument for ${\mathfrak {su}}(3)$ and
numerical investigations (such as that done for ${\mathfrak {su}}(5)$
in \cite{klei1}) for higher dimension groups 
which indicate that this
is in fact the case.
The method used to prove the main theorem of this paper is 
remarkably simple, drawing only on elementary topological ideas.

\bigskip
The structure of the paper is as follows. 
In section 2 we review briefly the ${\mathfrak {su}}(N)$
Einstein-Yang-Mills field equations and the ansatz and
notations we shall employ in the rest of the paper.
Next we state some elementary properties of these equations,
including results from \cite{kunz1}.
The remainder of the paper has a similar progression of ideas
as \cite{bfm}, and we continue by first analyzing the
behaviour of solutions to the field equations in the two asymptotic
regimes, at infinity and close to the event horizon.
Although these forms were discussed in \cite{kunz1}, we present
here a shorter proof.
Section 5 is devoted to a discussion of the flat space solutions,
which will be important for later propositions.  
The results here are somewhat weaker than in \cite{bfm} due
to the fact that for ${\mathfrak {su}}(N)$, we have 
$N-1$ variables and therefore have a $2(N-1)$-dimensional phase
space rather than a phase plane as in the ${\mathfrak {su}}(2)$
case, so the powerful Poincar\'e-Bendixson theory no longer applies.
We now consider integrating the field equations outward from 
the event horizon and consider the various possible behaviours
of the resulting solutions.
As in \cite{bfm}, there are three types of solution: the regular
black holes we are seeking, singular solutions (in which the
lapse function vanishes outside the event horizon), and 
oscillating solutions in which the geometry is not asymptotically 
flat.
The main results of this paper are in sections 7 and 8.
An inductive argument is used to prove the existence of solutions
for ${\mathfrak {su}}(N)$ assuming existence for 
${\mathfrak {su}}(N-1)$ (since we have rigorous theorems for
${\mathfrak {su}}(2)$ \cite{bfm,smoller}).
The argument is presented in detail for ${\mathfrak {su}}(3)$
in section 7 and a brief outline of the extension to general
$N$ in section 8.
Finally, a summary and our conclusions are presented in section 9.

\section{Ansatz and field equations}
\label{ansatz}
In this section we first describe the ansatz we are using and outline
the field equations.

\bigskip
The field equations for an ${\mathfrak {su}}(N)$ 
Yang-Mills gauge field coupled to
gravity have been derived in \cite{kunzle} for a spherically symmetric
geometry.
We take the line element, in the usual Schwarzschild co-ordinates,
to be
\be
ds^{2}=-  S^{2}\mu \, dt^{2} + \mu ^{-1}\, dr^{2}+
r^{2} \, d\theta ^{2} + r^{2} \sin ^{2} \theta \, d\phi ^{2},
\ee
where the metric functions $\mu $ and $S$ are functions of $r$ alone
and 
\be
\mu (r) = 1-\frac {2m(r)}{r}.
\ee
A spherically symmetric ${\mathfrak {su}}(N)$ 
gauge potential may be written in the
form \cite{kunzle}
\be
{\cal {A}}=A\,dt+B\, dr + 
\frac {1}{2} \left( C-C^{H} \right) \, d\theta 
-\frac {i}{2} \left[ \left( C+C^{H} \right) \sin \theta
+D\cos \theta \right] \, d\phi
\ee
where 
$D={\mbox {Diag}} \left\{ k_{1}, k_{2}, \ldots , k_{N} \right\} $
with $k_{1}\ge k_{2} \ge \ldots \ge k_{N}$ integers whose sum is zero.
In addition, $C$ is a strictly upper triangular complex matrix 
such that $C_{ij}\neq 0$ only if $k_{i}=k_{j}+2$ and $C^{H}$ its
Hermitian conjugate, and $A$, $B$ are anti-Hermitian matrices that
commute with $D$.
An irreducible representation of ${\mathfrak {su}}(N)$ 
can be constructed by taking
\cite{kunzle}
\be
D={\mbox {Diag}}\left\{ N-1, N-3, \ldots , -N+3, -N+1 \right\} .
\ee
In this case $A$, $B$ have trace zero and can be written as
\be
A_{jj}=i\left\{
-\frac {1}{N} \sum _{k=1}^{j-1} k{\cal {A}}_{k}
+\sum _{k=j}^{N-1} \left( 1-\frac {k}{N} \right) {\cal {A}}_{k}
\right\}
\ee
for real functions ${\cal {A}}_{k}$, and similarly for $B$.
The only non-vanishing entries of $C$ are
\be
C_{j,j+1}= \omega _{j} e^{i\gamma _{j}}
\ee
where $\omega _{j}$ and $\gamma _{j}$ are real functions.
All the functions in this ansatz depend only on $r$.
Now we make the following simplifying assumption \cite{kunzle}:
\be
{\cal {A}}_{j}=0, \qquad
{\cal {B}}_{j}+\gamma _{j}'=0,
\qquad
\forall j.
\ee
The remaining gauge freedom can then be used to set $B=0$
\cite{brod}, which implies that the $\gamma _{j}$ are constants, which
we choose to be zero for simplicity.

\bigskip
The gauge field equations for the $\omega _{j}$ then take the form
\cite{kunzle,kunz1}(where we have set $\kappa =2$ in \cite{kunzle})
\be
r^{2} \mu \omega _{j}'' + \left( 2m-2r^{3} p_{\theta } \right)
\omega _{j}' + 
\left[ 1-\omega _{j}^{2} + \frac {1}{2} \left(
\omega _{j-1}^{2} + \omega _{j+1}^{2} \right) \right] \omega _{j}
=0,
\label{firsteqn}
\ee
where
\be
p_{\theta }=\frac {1}{4r^{4}} \sum _{j=1}^{N}
\left[ \omega _{j}^{2} -\omega _{j-1}^{2} -N -1 +2j 
\right] ^{2}
\ee
and the Einstein equations can be simplified to
\be
m'= \mu G + r^{2} p_{\theta } ,
\qquad
\frac {S'}{S} =\frac {2G}{r}
\label{Seqn}
\ee
where
\be
G= \sum _{j=1}^{N-1} \omega _{j}^{'2}.
\label{lasteqn}
\ee
Note that $\omega _{0}\equiv 0 \equiv \omega _{N}$, so that these
are the usual equations \cite{bizon} in the ${\mathfrak {su}}(2)$ 
case, and have a 
very similar, but slightly coupled, structure for general $N$.
The equations (\ref{firsteqn}--\ref{lasteqn}) possess two symmetries
\cite{kunz1}.
Firstly, the substitution $\omega _{j}\rightarrow -\omega _{j}$
for any fixed $j$ leaves the equations invariant, exactly as in the
${\mathfrak {su}}(2)$ case.  
Secondly, there is an additional symmetry under the transformation
$j\rightarrow N-j$ for all $j$.
We will not assume that this symmetry is respected by the 
solutions of the field equations, so that the ${\mathfrak {su}}(3)$
case is not necessarily trivial.

\bigskip
We are concerned in this paper with black holes, which will have a
regular event horizon at $r=r_{h}$, where $\mu =0$.
The equations in their present form are singular at $r_{h}$, so in
order to produce a set of equations which are regular as 
$\mu \rightarrow 0$, we define a new independent variable $\tau $ 
by \cite{bfm}
\be
\frac {dr}{d\tau }= r {\sqrt {\mu }},
\ee
denote $d/d\tau $ by ${\dot {}}$, and define new dependent variables
$\kappa $, $U_{j}$ and $\Psi $ as follows \cite{bfm}:
\be
\Psi = {\sqrt {\mu }}, \qquad
U_{j}=\Psi \omega _{j}', \qquad
\kappa = \frac {1}{2\Psi } \left(
1+\Psi ^{2} + 2\mu G - 2r^{2} p_{\theta } \right) .
\ee
Then the field equations take the form \cite{bfm}:
\bea
{\dot {r}} & = & r\Psi 
\label{fregeqn} \\
{\dot {\omega }}_{j} & = & rU_{j} \\
{\dot {\Psi }} & = & 
	\left( \kappa - \Psi \right) \Psi - 2\mu G \\
\left( S\Psi \right) {\dot {}} & = & 
	S\left( \kappa - \Psi \right) \Psi \\
{\dot {U}}_{j} & = & 
	-\left( \kappa - \Psi \right) U_{j} 
	-\frac {1}{r} \left[ 
	1-\omega _{j}^{2} +\frac {1}{2} \left(
	\omega _{j+1}^{2} + \omega _{j-1}^{2} \right) \right] 
	\omega _{j} \\
{\dot {\kappa }} & = & 
	-\kappa ^{2} +1 +2\mu G.
\label{lregeqn}
\eea

\bigskip
The main thrust of this article is to prove the existence of regular,
black hole, solutions of the field equations for general $N$ with 
$N-1$ degrees of freedom. 
In other words, we begin with a regular event horizon at $r=r_{h}$,
where $\mu =0$, and integrate outwards with increasing $r$.
Later we shall classify the possible behaviour of the solutions 
as $r$ increases.
We are interested primarily in those solutions which possess the 
physical properties of a black hole geometry, namely for which
$\mu >0$ for all $r>r_{h}$, the $\omega _{j}$ and their derivatives
are finite for all $r>r_{h}$ and the spacetime becomes flat in the
limit  $r\rightarrow \infty $.
We shall refer to such solutions as
{\em {regular black hole solutions}}.

\section{Elementary results}
In this this section we state a few elementary results, which
will prove useful in later analysis.
Firstly, two lemmas which are proved in the
${\mathfrak {su}}(N)$ case exactly as in the ${\mathfrak {su}}(2)$
case, see \cite{bfm}.

\begin{lemma}
If $\mu (r_{0})<1$ for some $r_{0}$ then $\mu (r)<1$ for all 
$r\ge r_{0}$.
\end{lemma}

\begin{lemma}
\label{easylem}
As long as $0<\mu <1$ all field variables are regular functions
of $r$.
\end{lemma}

These two lemmas show that for a regular event horizon at 
$r=r_{h}$, where $\mu =0$, then $\mu <1$ for all $r>r_{h}$
and the field variables are regular functions as long as 
$\mu >0$.
The black hole solutions we seek approach the Schwarzschild geometry
as $r\rightarrow \infty $, so  we may
assume that all $\omega _{j}$ are bounded for all $r$ (from lemma
\ref{easylem} each $\omega _{j}$ is bounded on every closed interval
$[r_{h},r_{1}]$, so we are simply assuming that $\omega _{j}$
remains bounded as $r\rightarrow \infty $).
It will be proved in section \ref{global} that this 
assumption is in fact valid for solutions having $\mu >0$
for all $r$.
Define a quantity $M_{j}$ for each $j$ to be the lowest upper bound
on $\omega _{j}^{2}$, i.e.
\be
\omega _{j}^{2} \le M_{j}, \qquad \forall j=1,2,\ldots, N-1.
\ee
It has been shown in \cite{smoller} that the following result
is true for $N=2$, and proved in general with the assumption
that all $\omega _{j}$ are bounded via an elegant method in \cite{kunz1}.

\begin{theorem}
\label{mtheorem}
$M_{j}\le j(N-j)$.
\end{theorem}

Part of the proof of this theorem in \cite{kunz1} involves a 
knowledge of where $\omega _{j}$ may have maxima or minima.
The result following is less powerful than in the ${\mathfrak {su}}(2)$
case due to the coupling in the gauge field equation (\ref{firsteqn}).

\begin{prop}
\label{maxminprop}
As long as $\mu (r)>0$, the function $\omega _{j}(r)$ cannot have
maxima in the regions
\be
\omega _{j}>{\sqrt {1+\frac {1}{2}\left( 
	\omega _{j+1}^{2} +\omega _{j-1}^{2} \right) }}
\qquad {\mbox {and}}
\qquad
0>\omega _{j}> -{\sqrt {1+\frac {1}{2}\left( 
	\omega _{j+1}^{2} +\omega _{j-1}^{2} \right) }}
\ee
or minima in the regions
\be
\omega _{j}<-{\sqrt {1+\frac {1}{2}\left( 
	\omega _{j+1}^{2} +\omega _{j-1}^{2} \right) }}
\qquad {\mbox {and}}
\qquad
0<\omega _{j}<{\sqrt {1+\frac {1}{2}\left( 
	\omega _{j+1}^{2} +\omega _{j-1}^{2} \right) }}.
\ee
\end{prop}

\proof
When $\omega _{j}'=0$, from (\ref{firsteqn}),
\be
\mu \omega _{j}''=-\frac {1}{r^{2}} \left[
1-\omega _{j}^{2} + \frac {1}{2} \left(
\omega _{j+1}^{2} +\omega _{j-1}^{2} \right) \right]
\omega _{j},
\ee
so that $\omega _{j}$ will have a maximum if
\be
\omega _{j}>0 \qquad {\mbox {and}} \qquad
\omega _{j}^{2}<1+\frac {1}{2} \left(
\omega _{j+1}^{2} +\omega _{j-1}^{2} \right)
\ee
or
\be
\omega _{j}<0 \qquad {\mbox {and}} \qquad
\omega _{j}^{2}>1+\frac {1}{2} \left(
\omega _{j+1}^{2} +\omega _{j-1}^{2} \right) .
\ee
Similarly, $\omega _{j}$ will have a minimum if
\be
\omega _{j}>0 \qquad {\mbox {and}} \qquad
\omega _{j}^{2}>1+\frac {1}{2} \left(
\omega _{j+1}^{2} +\omega _{j-1}^{2} \right) 
\ee
or
\be
\omega _{j}<0 \qquad {\mbox {and}} \qquad
\omega _{j}^{2}<1+\frac {1}{2} \left(
\omega _{j+1}^{2} +\omega _{j-1}^{2} \right) .
\ee
\hfill
$\square $

\section{Asymptotic behaviour of solutions}
\label{local}
We seek solutions to the above field equations representing black
holes with a regular event horizon at $r=r_{h}$ and finite total
energy density. 
In order to prove the local existence of solutions of the field
equations with the desired asymptotic behaviour, 
we shall apply the following
theorem \cite{bfm}.

\begin{theorem}
\label{bfmth}
Consider a system of differential equations for $n+m$ functions
${\mbox {\boldmath {$a$}}}=(a_{1}, a_{2}, \ldots , a_{n})$
and
${\mbox {\boldmath {$b$}}}=(b_{1}, b_{2}, \ldots , b_{m})$
of the form
\bea
x \frac {da_{i}}{dx} & = & 
x^{p_{i}} f_{i}(x,{\mbox {\boldmath {$a$}}},{\mbox {\boldmath {$b$}}}),
\nonumber \\
x\frac {db_{i}}{dx} & = &
-\lambda _{i} b_{i} + 
x^{q_{i}} g_{i}(x,{\mbox {\boldmath {$a$}}},{\mbox {\boldmath {$b$}}}),
\label{sysde}
\eea
with constants $\lambda _{i}>0$ and integers $p_{i},q_{i}\ge 1$
and let ${\cal {C}}$ be an open subset of ${\mathbb {R}}^{n}$
such that the functions $f_{i}$ and $g_{i}$ are analytic in a
neighbourhood of $x=0$, 
${\mbox {\boldmath {$a$}}}={\mbox {\boldmath {$c$}}}$,
${\mbox {\boldmath {$b$}}}={\mbox {\boldmath {$0$}}}$, for all
${\mbox {\boldmath {$c$}}}\in {\cal {C}}$.
Then there exists an $n$-parameter family of solutions of the
system (\ref{sysde}) such that
\be
a_{i}(x) = c_{i} + O(x^{p_{i}}), \qquad
b_{i}(x) = O(x^{q_{i}}),
\ee
where $a_{i}(x)$ and $b_{i}(x)$ are defined for 
${\mbox {\boldmath {$c$}}}\in {\cal {C}}$, and 
$|x|<x_{0}({\mbox {\boldmath {$c$}}})$ and are analytic in $x$ and
${\mbox {\boldmath {$c$}}}$.
\end{theorem}
In this section we consider only the equations for $\mu $ and
the $\omega _{j}$. 
The equation for $S$ (\ref{Seqn}) is independent of the others and
hence can be integrated immediately once we have the other field 
variables, with an additional parameter. 
This procedure gives a finite answer on any closed, bounded interval
on which the remaining variables are finite.
Results similar to propositions \ref{infprop} and \ref{horprop}
were proved in \cite{kunz1}, but here we are using a more
straightforward approach.

\subsection{Behaviour at infinity}
As $r\rightarrow \infty$, in order to have finite total energy
density, the geometry must approach the Schwarzschild solution, that
is
\be
m\rightarrow M={\mbox {constant}}, \qquad
G\rightarrow 0, \qquad
p_{\theta } \rightarrow 0.
\label{inflimits}
\ee
Hence each $\omega _{j}$ must approach a constant value, which is
fixed by (\ref{inflimits}) to be
\be
\omega _{j}^{2} \rightarrow j(N-j).
\ee
The equations for the $\omega _{j}$ are then automatically satisfied
in the limit as $r\rightarrow \infty $.

\begin{prop}
\label{infprop}
There exists an $N$-parameter family of local solutions of 
(\ref{firsteqn}--\ref{lasteqn}) near $r=\infty $ analytic in $c_{j}$, 
$M$ and $r^{-1}$ such that
\bea
\mu (r) & = & 1-\frac {2M}{r} +O\left( \frac {1}{r^{4}} \right)
\nonumber \\
\omega _{j}(r) & = &  {\sqrt {j(N-j)}} - 
	\frac {c_{j}}{r} + O\left( \frac {1}{r^{2}} \right) .
\label{infrels}
\eea
\end{prop}

\proof
Introduce new variables
\be
x=\frac {1}{r}, \qquad
\psi _{j}=r\left( {\sqrt {j(N-j)}} -\omega _{j} \right), \qquad
\xi _{j}= r^{2} \mu \omega _{j}' \qquad
\lambda = r\left( 1-\mu \right) .
\ee
Then 
\be
x\frac {d\lambda }{dx} = -\frac {2}{x} \left( 
	\mu G +r^{2}p_{\theta } \right)
\ee
where 
\be
G=x^{4} \sum _{j=1}^{N-1} \frac {\xi _{j}^{2}}{\mu ^{2}}
\ee
and
\bea
p_{\theta } & = & \frac {x^{4}}{4} \sum _{j=1}^{N} \left[
x^{2}\psi _{j}^{2} - x^{2} \psi _{j-1}^{2} 
-2x\psi _{j}{\sqrt {j(N-j)}}
\right. \nonumber \\
& & \left.
+2x\psi _{j-1}{\sqrt {(j-1)(N-j+1)}} \right] ^{2},
\eea
that is,
\be
x\frac {d\lambda }{dx}=-x^{3}f_{\lambda }
\label{lambdaeqn}
\ee
where $f_{\lambda }$ is a polynomial in $x$, $\mu ^{-1}$, $\xi $'s and
$\psi $'s.
Similarly,
\be
x\frac {d\psi _{j}}{dx} = -\psi _{j}+\frac {1}{\mu }\xi _{j}
=-\psi _{j}+\xi _{j}+xf_{\psi _{j}}
\ee
where $f_{\psi _{j}}$ is analytic in $x$, $\lambda $, $\psi $'s
and $\xi $'s in a neighbourhood of $x=0$.
Using the field equations we also have
\bea
x\frac {d\xi _{j}}{dx} & = &
-2\xi _{j} +2 \psi _{j}\left[ j(N-j)\right]
-\psi _{j-1} {\sqrt {j(j-1)(N-j)(N-j+1)}}
\nonumber \\
& &
-\psi _{j+1} {\sqrt {j(j+1)(N-j)(N-j-1)}} 
+xf_{\xi _{j}}
\eea
where $f_{\xi _{j}}$ is a polynomial in $x$, $\mu ^{-1}$, $\psi $'s
and $\xi $'s.
The algebra simplifies if we
define new functions $\alpha _{j}$, $\beta _{j}$ such that
\bea
\psi _{j} & = & \alpha _{j}+\beta _{j}
\nonumber
\\
\xi _{j} & = & \alpha _{j}+\Lambda \beta _{j}
\eea
where  $\Lambda $ is a real constant not equal to unity.
The equations for $\alpha _{j}$ and $\beta _{j}$ then read
\bea
(\Lambda -1)x\frac {d\alpha _{j}}{dx} & = &  
2\alpha _{j}-2[j(N-j)]\alpha _{j} +
{\sqrt {j(j-1)(N-j)(N-j+1)}}\alpha _{j-1} 
\nonumber \\
& &
+
{\sqrt {j(j+1)(N-j)(N-j-1)}}\alpha _{j+1} +
(\Lambda^{2}+\Lambda)\beta _{j} 
\nonumber
\\
 & &
-
2[j(N-j)]\beta _{j} +
{\sqrt {j(j-1)(N-j)(N-j+1)}}\beta _{j-1} 
\nonumber 
\\
& & 
+
{\sqrt {j(j+1)(N-j)(N-j-1)}}\beta _{j+1} +
xf_{\alpha _{j}}
\nonumber
\\
(\Lambda-1)x\frac {d\beta _{j}}{dx} & = & 
-2\alpha _{j}+2[j(N-j)]\alpha _{j} -
{\sqrt {j(j-1)(N-j)(N-j+1)}}\alpha _{j-1} 
\nonumber 
\\ 
& &
-
{\sqrt {j(j+1)(N-j)(N-j-1)}}\alpha _{j+1} +
(1-3\Lambda)\beta _{j} 
\nonumber 
\\
 & & 
+
2[j(N-j)]\beta _{j} -
{\sqrt {j(j-1)(N-j)(N-j+1)}}\beta _{j-1} 
\nonumber
\\
& &
-
{\sqrt {j(j+1)(N-j)(N-j-1)}}\beta _{j+1} +
xf_{\beta _{j}} ,
\label{alphabeta}
\eea
where the $f_{\alpha _{j}}$ and $f_{\beta _{j}}$ are analytic
functions of $x$, $\mu ^{-1}$, $\alpha $'s and $\beta $'s.
Consider the matrix ${\cal {M}}_{N-1}$ whose entries are
\be
\left(
\begin{array}{ccc}
2(N-1) & -{\sqrt {(N-1)2(N-2)}} & \ldots  \\
-{\sqrt {(N-1)2(N-2)}} & 2.2(N-2) & \ldots \\
0 & -{\sqrt {2(N-2)3(N-3)}} & \ldots \\
\vdots & \vdots & \vdots  \\ 
\end{array}
\right) .
\label{matrix}
\ee  
For a vector ${\mbox {\boldmath {$q$}}}=\left( 
q_{1},\ldots , q_{N-1} \right) $,  let 
$p_{j}={\sqrt {j(N-j)}}q_{j}$ for $j=1, \ldots , N-1$.
Then
\be
{\mbox {\boldmath {$q$}}}^{T} {\cal {M}}_{N-1} 
{\mbox {\boldmath {$q$}}}
=p_{1}^{2}+p_{N-1}^{2}+
\left( p_{1}-p_{2} \right) ^{2} +\ldots
\left( p_{N-2} -p_{N-1} \right) ^{2} .
\ee
Hence the matrix ${\cal {M}}_{N-1}$ is real, symmetric and positive
definite, and will have positive real eigenvalues ${\cal {E}} _{i}$,
and corresponding eigenvectors ${\mbox {\boldmath {$v$}}}_{i}$.
We may expand our variable vectors in terms of eigenvectors
of ${\cal {M}}_{N-1}$ as follows:
\be
{\mbox {\boldmath {$\alpha $}}}=
\sum _{i=1}^{N-1}A_{i}(x){\mbox {\boldmath {$v$}}}_{i}
\qquad
{\mbox {\boldmath {$\beta $}}}=
\sum _{i=1}^{N-1}B_{i}(x){\mbox {\boldmath {$v$}}}_{i}
\label{subs1}
\ee
\be
{\mbox {\boldmath {$f$}}}_{\alpha }=
\sum _{i=1}^{N-1}F_{i}(x){\mbox {\boldmath {$v$}}}_{i}
\qquad
{\mbox {\boldmath {$f$}}}_{\beta }=
\sum _{i=1}^{N-1}G_{i}(x){\mbox {\boldmath {$v$}}}_{i},
\ee
in terms of which the equations (\ref{alphabeta}) now become:
\bea
(\Lambda -1)x\frac {dA_{i}}{dx} & = & 
(2-{\cal {E}}_{i})A_{i}+
(\Lambda ^{2}+\Lambda -{\cal {E}}_{i})B_{i}+
xF_{i}
\nonumber
\\
(\Lambda -1)x\frac {dB_{i}}{dx} & = & 
({\cal {E}}_{i}-2)A_{i}+
(1-3\Lambda +{\cal {E}}_{i})B_{i}+xG_{i}.
\eea
This transformation has removed the coupling between the $j$'s but
the $A$'s and $B$'s are still coupled.
Repeating the procedure with the matrix ${\cal {P}}_{i}$ will 
decouple these equations, where
\be
{\cal {P}}_{i}=\left(
\begin{array}{cc}
2-{\cal {E}}_{i} & \Lambda ^{2}+\Lambda -{\cal {E}}_{i} \\
-2+{\cal {E}}_{i} & 1-3\Lambda +{\cal {E}}_{i} 
\end{array}
\right) .
\ee
This matrix has eigenvalues
\be
\mu _{\pm }=
\frac {1}{2} (1-\Lambda ) \left(
3\pm {\sqrt {1+4{\cal {E}}_{i}}} \right)
\ee
and corresponding eigenvectors ${\mbox {\boldmath {$V$}}}_{\pm }$.
Writing
\be
\left(
\begin{array}{c}
A_{i} \\ B_{i} 
\end{array}
\right)
= C_{+}(x){\mbox {\boldmath {$V$}}}_{+}
+C_{-}(x){\mbox {\boldmath {$V$}}}_{-},
\qquad
\left(
\begin{array}{c}
F_{i} \\ G_{i} 
\end{array}
\right)
= D_{+}(x){\mbox {\boldmath {$V$}}}_{+}
+D_{-}(x){\mbox {\boldmath {$V$}}}_{-},
\label{subs2}
\ee
the equations are now
\be
(\Lambda -1)x\frac {dC_{\pm }}{dx} =
\mu _{\pm }C_{\pm }+xD_{\pm }.
\label{eigeneqns}
\ee
In the case where $\mu _{-}<0$, theorem \ref{bfmth} applies directly
and we have
\be
C_{-}=O(x);
\ee
when $\mu _{-}=0$ (corresponding to ${\cal {E}}_{i}=2$), 
\be
C_{-}=K_{-}+O(x)
\ee
for some constant $K_{-}$.
The other eigenvalue $\mu _{+}$ is always strictly positive, and 
in this situation  theorem \ref{bfmth} does not apply.
However, equation (\ref{eigeneqns}) can be integrated using the
standard Picard method (see, for example, \cite{codd}) to
give an analytic solution
\be
C_{+}=K_{+}+O(x).
\ee
Both $C_{\pm }$ will be analytic in $K_{\pm }$ and $x$ in some
neighbourhood of $x=0$ and substituting back through equations
(\ref{subs1},\ref{subs2}) gives solutions of the form
(\ref{infrels}) as required.
\hfill
$\square $

\subsection{Behaviour at the event horizon}
We assume that at $r=r_{h}$ there is a non-degenerate event horizon,
namely $\mu (r_{h})=0$, but $\mu '(r_{h})>0$ is finite.

\begin{prop}
\label{horprop}
There exists an $N$-parameter family of local solutions of
(\ref{firsteqn}--\ref{lasteqn}) near $r=r_{h}$, analytic in $r_{h}$,
$\omega _{j,h}$ and $r$ such that
\bea
\mu (r_{h}+\rho ) & = & \mu '(r_{h})+O(\rho ), 
\nonumber \\
\omega _{j} (r_{h}+\rho ) & = & 
	\omega _{j,h} +\omega _{j}'(r_{h}) +O(\rho ^{2})
\label{horrels}
\eea
where
$\mu '(r_{h})$ and $\omega _{j}'(r_{h})$ are functions of the
$\omega _{j,h}$.
\end{prop}

\proof
Let $\rho =r-r_{h}$ be the new independent variable, and define
\be
x=r, \qquad
\lambda = \frac {\mu }{\rho }, \qquad
\psi _{j}=\omega _{j}, \qquad
\xi _{j}=\frac {\mu \omega _{j}'}{\rho }.
\ee
Then the field equations take the form
\bea
\rho \frac {dx}{d\rho } & = & \rho, \\
\rho \frac {d\lambda }{d\rho } & = & 
	-\lambda + \left[ \frac {1}{x} - 2xp_{\theta }\right]
	+\rho \frac {\lambda }{x}\left[ 1-2G \right] 
\nonumber \\
 & = & 
	-\lambda +\rho H_{\lambda }+F_{\lambda } \\
\rho \frac {d\psi _{j}}{d\rho } & = & 
	\frac {\rho \xi _{j}}{\lambda } \\
\rho \frac {d\xi _{j}}{d\rho } & = & 
	-\xi _{j} +\rho H_{j} + F_{j} ,
\eea
where the $F$'s and $H$'s are polynomials in $x^{-1}$, 
$\lambda ^{-1}$, and the other variables, and
the $F$'s depend only on $x$ and $\psi _{j}$'s.
Next define
\be
{\tilde {\xi }}_{j} = \xi _{j}-F_{j}, \qquad
{\tilde {\lambda }} = \lambda -F_{\lambda },
\ee
whose derivatives are given by
\bea
\rho \frac {d{\tilde {\xi }}_{j}}{d\rho } & = & 
	-{\tilde {\xi }}_{j}+\rho G_{j} \\
\rho \frac {d{\tilde {\lambda }}}{d\rho } & = & 
	-{\tilde {\lambda }} + \rho G_{\lambda }.
\eea
Here the $G$'s are analytic in $x^{-1}$, $\lambda ^{-1}$, $\lambda $,
$x$, ${\tilde {\lambda }}$, $\psi _{j}$, ${\tilde {\xi }}_{j}$.
Applying theorem \ref{bfmth}, there exist solutions of the form
\be
x=r_{h}+\rho , \qquad
\psi _{j}=\omega _{j,h}+O(\rho ), \qquad
{\tilde {\lambda }},{\tilde {\xi }}_{j}=O(\rho ),
\ee
which gives the behaviour (\ref{horrels}), together with the
required analyticity.
From the field equations (\ref{firsteqn}--\ref{lasteqn}), setting
$\mu (r_{h})=0$ gives the following relations:
\bea
\mu '(r_{h}) & = &  \frac {1}{2} r_{h}^{2} p_{\theta }(r_{h}) \\
\omega _{j}'(r_{h}) & = & -
\frac {\left[ 1-\omega _{j,h}^{2}+\frac {1}{2} \left(
\omega _{j-1,h}^{2}+\omega _{j+1,h}^{2} \right) \omega _{j,h}
\right] }{r_{h} -r_{h}^{3}p_{\theta }(r_{h})} 
\eea
where
\be
p_{\theta }(r_{h})=
\frac {1}{4r_{h}^{4}} \sum _{j=1}^{N} \left[
\omega _{j,h}^{2} -\omega _{j-1,h}^{2} -N-1+2j \right] ^{2}.
\ee
\hfill
$\square $

\section{Flat space solutions}
\label{flat}
The behaviour of the gauge field equations in the flat space limit
will be useful later in section \ref{global} when we come to examine the 
properties of regular solutions.
We are not concerned here with the global existence of flat space 
solutions but only the local properties pertinent to the curved space
problem.

\bigskip
The analysis here is similar to that in \cite{bfm}, but note that
for general $N$ there will be $N-1$ coupled gauge field equations.
The powerful Poincar\'e-Bendixson theory of autonomous systems 
is not applicable when $N>2$, and so we will be able to derive 
only correspondingly weaker results, and cannot draw a phase portrait.
Fortunately the theorems later require only a knowledge of the 
nature of the critical points and the local behaviour close to these
points, which is the subject of this section.

\bigskip
In flat space the gauge field equations reduce to:
\be
r^{2} \omega _{j}'' +\left[ 1-\omega _{j}^{2} +\frac {1}{2}
\left( \omega _{j+1}^{2} +\omega _{j-1}^{2} \right) \right] 
\omega _{j} =0.
\ee
These equations can be made autonomous by changing variables to
$\tau =\log r$, and denoting $d/d\tau $ by ${\dot {}}$:
\be
{\ddot {\omega }}_{j}-{\dot {\omega }}_{j}+
\left[ 1-\omega _{j}^{2} +\frac {1}{2}
\left( \omega _{j+1}^{2} +\omega _{j-1}^{2} \right) \right] 
\omega _{j} =0.
\ee
This system of $N-1$ coupled equations has critical points when
\be
\left[ 1-\omega _{j}^{2} +\frac {1}{2}
\left( \omega _{j+1}^{2} +\omega _{j-1}^{2} \right) \right] 
\omega _{j} =0.
\qquad
\forall j=1,2,\ldots ,N-1.
\ee
If $\omega _{j}\neq 0$ for all $j$ then 
\be
\omega _{j}={\sqrt {j(N-j)}}.
\ee
There is also a critical point when $\omega _{j}=0$ for all $j$.

\bigskip
The other critical points can be described as follows.
Suppose that $\omega _{i}=0$ and $\omega _{i+k}=0$ but that
$\omega _{i+m}\neq 0$ for $m=1,\ldots ,k-1$ (where $k\ge 2$).
The critical point is then described by the equations
\bea
0 & = & \omega _{i} \nonumber \\
0 & = & 1-\omega _{i+1}^{2}+\frac {1}{2} \left( \omega _{i}^{2}
+\omega _{i+2}^{2} \right) 
\nonumber 
\\
 & \vdots & 
\nonumber
\\
0 & = & 1-\omega _{i+k-1}^{2} +\frac {1}{2} \left( \omega _{i+k-2}^{2}
+\omega _{i+k}^{2} \right)
\nonumber
\\
0 & = & \omega _{i+k}
\eea
which have the solution
\be
\omega _{i+m}={\sqrt {m(k-m)}} \qquad 
{\mbox {for $m=1,\ldots , k-1$}}.
\ee
In other words, if we have a run of $k$ non-zero $\omega $'s with
zero $\omega _{i}$ at each end, then the solution for the non-zero
$\omega $'s is the same as the $N=k$ case with no zero $\omega $'s.
We can also put together a run of as many zero $\omega $'s as we like
in the critical point.

\bigskip
In order to classify the critical points, we linearize the field
equations.
Let
\be
\omega _{j}(\tau )=\omega _{j}^{(0)}+\epsilon _{j}(\tau )
\ee
where $\omega _{j}^{(0)}$ is the value of $\omega _{j}$ at
the critical point and $\epsilon _{j}$ is a small perturbation.
The equation for $\epsilon _{j}$ is, to first order,
\bea
0 & = & 
{\ddot {\epsilon }}_{j} -{\dot {\epsilon }}_{j} 
+\epsilon _{j}\left[ 1-\omega _{j}^{(0)2} +\frac {1}{2}
\left( \omega _{j+1}^{(0)2} +\omega _{j-1}^{(0)2} \right) \right] 
\nonumber
\\ 
 &  & 
+\omega _{j}^{(0)} \left[ -2\epsilon _{j}\omega _{j}^{(0)}
+\epsilon _{j+1} \omega _{j+1}^{(0)} +\epsilon _{j-1}
\omega _{j-1}^{(0)} \right] .
\eea
We shall take  the two cases we need to consider in turn.

\bigskip
Firstly, the case where $\omega _{j}^{(0)}=0$, when the equation
for $\epsilon _{j}$ reduces to
\be
{\ddot {\epsilon }}_{j}-{\dot {\epsilon }}_{j}+
\epsilon _{j} \left[ 1+\frac {1}{2} \left(
\omega _{j+1}^{(0)2} +\omega _{j-1}^{(0)2} \right) \right] =0.
\ee
In order to find the nature of the critical point, let
\be
\epsilon _{j}=e^{\lambda \tau },
\ee
then $\lambda $ satisfies the equation
\be
\lambda ^{2}-\lambda +\left[ 
1+\frac {1}{2} \left(
\omega _{j+1}^{(0)2} +\omega _{j-1}^{(0)2} \right) \right] =0.
\ee
This implies that
\be
\lambda = \frac {1}{2} \pm \frac {1}{2} i\alpha,
\ee
where $\alpha $ is the  positive real number given by
\be
\alpha ^{2} = 3+2\omega _{j+1}^{(0)2} +2\omega _{j-1}^{(0)2}.
\ee
We conclude that, in the $(\epsilon _{j}, {\dot {\epsilon }}_{j})$
plane, we have an unstable focal point, as found in the ${\mathfrak {su}}(2)$ 
case in \cite{bfm}.

\bigskip
Secondly, we consider the situation in which there is at least
one $\omega _{j}^{(0)}$ which is non-zero.
Suppose that $\omega _{i}^{(0)}=0$, and $\omega _{i+k}^{(0)}=0$,
but  $\omega _{i+m}^{(0)}\neq 0$ for $m=1,\ldots ,k-1$, where we 
include the case that $i=0$ and $k=N$.
Then we have a series of coupled perturbation equations:
\bea
0 & = & 
{\ddot {\epsilon }}_{i+1} - {\dot {\epsilon }}_{i+1}
-2\epsilon _{i+1} \omega _{i+1}^{(0)2} 
+\epsilon _{i+2} \omega _{i+2}^{(0)} \omega _{i+1}^{(0)} 
\nonumber \\
0 & = & 
{\ddot {\epsilon }}_{i+2} - {\dot {\epsilon }}_{i+2}
-2\epsilon _{i+2} \omega _{i+2}^{(0)2} 
+\epsilon _{i+3} \omega _{i+3}^{(0)} \omega _{i+2}^{(0)}
+\epsilon _{i+1} \omega _{i+1}^{(0)} \omega _{i+2}^{(0)}
\nonumber \\
 & \vdots & 
\nonumber \\
0 & = & 
{\ddot {\epsilon }}_{i+k-1} - {\dot {\epsilon }}_{i+k-1}
-2\epsilon _{i+k-1} \omega _{i+k-1}^{(0)2} 
+\epsilon _{i+k-2} \omega _{i+k-2}^{(0)} \omega _{i+k-1}^{(0)} .
\eea
Define a vector ${\mbox {\boldmath {$\epsilon $}}}$ by
\be
{\mbox {\boldmath {$\epsilon $}}} = \left( 
\epsilon _{i+1}, \ldots , \epsilon _{i+k-1} \right) ^{T}
\ee
and consider solutions of the form 
\be
{\mbox {\boldmath {$\epsilon $}}} = e^{\lambda \tau }
{\mbox {\boldmath {$q$}}}
\ee
where ${\mbox {\boldmath {$q$}}}$ is a constant vector.
Then $\lambda ^{2}-\lambda $ are eigenvalues of the matrix
${\cal {M}}_{k-1}$ given by (\ref{matrix}).
As discussed in section \ref{local}, the matrix ${\cal {M}}_{k-1}$ 
has positive real eigenvalues ${\cal {E}}_{j}$.
Then
\be
\lambda = \frac {1\pm {\sqrt {1+4{\cal {E}}_{j}}}}{2}
=j+1, -j
\ee
will have positive and negative real values and there is a 
saddle point.
Again this is in direct analogy with the 
${\mathfrak {su}}(2)$ case \cite{bfm}.

\section{Global behaviour of the solutions}
\label{global}
In this section we investigate the behaviour of solutions of
the regular field equations (\ref{fregeqn}--\ref{lregeqn}) as 
functions of $\tau $.
From section \ref{local}, we know that, given any starting values
$\omega _{j,h}$ for the gauge field functions, then there is 
a local solution of the field equations in a neighbourhood of 
the regular event horizon at $r=r_{h}$, $\tau =0$, which is analytic
in $\tau $ and the initial parameters.
Furthermore, from lemma \ref{easylem} as long as $0<\Psi <1$ the
solutions remain regular.
Therefore, as we integrate out in $\tau $ from the event horizon there
are only three possibilities:
\begin{enumerate}
\item
There is a $\tau _{0}>0$ such that $\Psi (\tau _{0})=0$.
\item
For all $\tau >0$, we have $\Psi (\tau )>0$ and $r(\tau )$ remains
bounded as $\tau \rightarrow \infty $.
\item
For all $\tau >0$, we have $\Psi (\tau )>0$ and 
$r(\tau )\rightarrow \infty $ as $\tau \rightarrow \infty $.
\end{enumerate}
We refer to solutions of the first type as {\em {singular}} solutions.
Those of type 2 are known as {\em {oscillating}} solutions in
\cite{bfm} and we retain their terminology.
It will be shown below (proposition \ref{rmaxprop}), that
there is a maximum value $r_{max}$ of $r$ for
oscillating solutions.
Finally, we denote by $S_{\infty }$ solutions of the
first type for which $r(\tau _{0})<r_{max}$. 
Our first task is to show that solutions of type 3 are precisely 
the black hole solutions we are seeking.
We begin by proving the assumption made at the end of
section \ref{ansatz},
namely that for solutions of type 3, each $\omega _{j}$ is
bounded for all $r$.

\begin{lemma}
\label{boundlem}
If $\Psi (\tau )>0$ for all $\tau $ and 
$\lim _{\tau \rightarrow \infty }r(\tau )=\infty $ then
all the $\omega _{j}$ remain bounded as $r\rightarrow \infty $.
\end{lemma}

\proof
From the local existence propositions \ref{horprop} and \ref{infprop}
and lemma \ref{easylem}, each $\omega _{j}$ is an analytic
function of $r$ as long as $\mu >0$.
Introduce a new variable $x=r^{-1}$, then each $\omega _{j}$
is an analytic function of $x$ in a neighbourhood of $x=0$,
except possibly at $x=0$, and can be written as a Laurent series:
\be
\omega _{j}(x)=\sum _{n=-\infty }^{\infty } a_{n}^{j}x^{n}.
\ee
In order for $\mu >0$ for all $r$, it must be the case that
$2mr^{-1}=2mx<1$, although $m$ itself need not necessarily 
remain bounded.
Hence $2mx$ is bounded in a neighbourhood of $x=0$, and in 
particular is analytic at $x=0$.
Therefore we can write
\be
2mx=\sum _{n=0}^{\infty }b_{n}x^{n}.
\ee
Thus
\be
m=\frac {1}{2} \sum _{n=0}^{\infty }b_{n}x^{n-1},
\qquad
\frac {dm}{dr}=-x^{2}\frac {dm}{dx}
=-\frac {1}{2} \sum _{n=0}^{\infty }b_{n}(n-1)x^{n},
\ee
so that $dm/dr$ is analytic in a neighbourhood of $x=0$.
From the field equations,
\be
\frac {dm}{dr}=\mu G +r^{2}p_{\theta }
\ee
where
\be
G=\sum _{j=1}^{N-1} \left( \frac {d\omega _{j}}{dr} \right) ^{2}
\qquad
r^{2}p_{\theta }=\frac {x^{2}}{4}
\sum _{j=1}^{N} \left[ 
\omega _{j}^{2}-\omega _{j-1}^{2}-N-1+2j \right] ^{2}.
\ee
Since both $G$ and $r^{2}p_{\theta }$ are positive, they must
each be analytic in a neighbourhood of $x=0$.
Consider $G$ first, using
\be
\frac {d\omega _{j}}{dr}=-x^{2}\frac {d\omega _{j}}{dx}
=-\sum _{n=-\infty }^{\infty }a_{n}^{j} nx^{n+1}
\ee
and since $\left( \frac {d\omega _{j}}{dr}\right) ^{2}$
must be analytic near $x=0$, it must be the case that
\be
a_{n}^{j}=0 \qquad \forall n<-1.
\ee
Then
\be
\omega _{j}^{2}=\frac {a_{-1}^{j2}}{x^{2}}+
\sum _{n=-1}^{\infty }c_{n}^{j}x^{n}
\ee
for constants $c_{n}^{j}$ and hence
\be
\omega _{j}^{2}-\omega _{j-1}^{2}-N-1+2j
=\frac {1}{x^{2}}\left[ a_{-1}^{j2}-a_{-1}^{(j-1)2} \right]
+\sum _{n=-1}^{\infty }d_{n}^{j}x^{n}.
\ee
Since $r^{2}p_{\theta }$ is analytic, 
\be
a_{-1}^{j}=\pm a_{-1}^{j-1}
\qquad
\forall j.
\ee
But $\omega _{0}\equiv 0$, so $a_{-1}^{0}=0$ and thus $a_{-1}^{j}=0$
for all $j$.
We conclude that $\omega _{j}$ is analytic at $x=0$, and therefore
finite as $x=0$.
Therefore $\omega _{j}$ is bounded for all $r\in [r_{h},\infty )$.
\hfill
$\square $

\begin{prop}
\label{flatprop}
If $\Psi (\tau )>0$ for all $\tau $ and 
$\lim _{\tau \rightarrow \infty }r(\tau )=\infty $ then
the solution tends to one of the flat space critical points, with 
the exception of the origin.
\end{prop}

The proof of this proposition closely follows 
Proposition 14 of Ref. 7, 
the main
thrust of which is the following lemma.

\begin{lemma}
\label{flatlem}
If $\Psi (\tau )>0$ for all $\tau $ and 
$\lim _{\tau \rightarrow \infty }r(\tau )=\infty $ then
\be
\lim _{\tau \rightarrow \infty } \Psi (\tau )=1.
\ee
\end{lemma}

\proof
There are two cases to consider.
\begin{enumerate}
\item
There is at least one $\omega _{j}$ which has zeros for arbitrarily
large $r$.
\item
All $\omega _{j}$ have only a finite number of zeros.
\end{enumerate}

\bigskip
\noindent
{\em {Case 1}}

\smallskip
\noindent
The proof in this situation follows exactly that of 
Proposition 14 of Ref. 7 (equations (74--76)) 
since all the $\omega _{j}$'s are 
bounded (lemma \ref{boundlem}).

\bigskip
\noindent
{\em {Case 2}}

\smallskip
\noindent
In this case there is a $T_{1}>0$ such that for all $\tau >T_{1}$
no $\omega _{j}$ has a zero.
However, since the equations governing the $\omega _{j}$ are
coupled, the $\omega _{j}$ do not necessarily have to be
monotonic, provided proposition \ref{maxminprop} is satisfied.

\bigskip
Suppose, initially, that there is a $T_{2}>T_{1}$ such that for all
$\tau >T_{2}$ every $\omega _{j}$ is monotonic.
In this situation each $\omega _{j}$ has a limit and so 
$\lim _{\tau \rightarrow \infty }U_{j}=0$ for all $j$.
Then the proof of Proposition 14 of Ref. 7 carries over directly
to show that $m(\tau )$ is bounded and 
$\lim _{\tau \rightarrow \infty } \Psi (\tau )=1$.
The proof proceeds as follows.
Firstly, in analogy to equation (74) of Ref. 7, integrating the
field equations (\ref{fregeqn}--\ref{lregeqn}) gives,
for each $j$,
\be
|\Psi U_{j}(\tau _{2})-\Psi U_{j}(\tau _{1})|
\le \frac {c_{j}}{r(\tau _{1})}
\ee
for all $\tau _{2}>\tau _{1}>T_{2}$, and some constant $c_{j}$,
since all the $\omega _{j}$ are bounded.
Fix $\tau _{1}$ for the moment, then for all $\tau _{3}>\tau _{1}$,
\be
\int _{\tau _{1}}^{\tau _{3}} r\Psi U_{j}^{2} \, d\tau '
\le c_{j}'\int _{\tau _{1}}^{\tau _{3}} {\dot {\omega }}_{j} \, 
d\tau ' \le
c_{j}''
\ee
for constants $c_{j}'$ and $c_{j}''$, since $\omega _{j}$ has a limit.
Hence
\be
m(\tau _{3})-m(\tau _{1})\le \frac {1}{2} \sum _{j=1}^{N-1}
c_{j}''
\ee
which is bounded.

\bigskip
Next turn to the other extreme situation, where $\omega _{j}^{2}$
has minima for arbitrarily large $r$, for every $j$.
If $\omega _{j}^{2}$ has a minimum at $r=r_{0}$, then by
proposition \ref{maxminprop},
\be
\omega _{j}^{2}(r_{0}) \ge 1+\frac{1}{2} \left(
\omega _{j+1}^{2} (r_{0}) +\omega _{j-1}^{2} (r_{0}) \right).
\ee
Define $N_{j}$ to be the greatest lower bound of the set of
minimum values of $\omega _{j}^{2}$ for $\tau >T_{1}$,
where $N_{j}$ can be zero (although $\omega _{j}^{2}$ cannot). 
Then we have
\be
N_{j}\ge 1+\frac {1}{2}\left(  N_{j+1}+N_{j-1} \right) .
\label{nineq}
\ee 
These inequalities may be solved to give
\be
N_{j}\ge j(N-j) \qquad \forall j.
\ee
However, since $\omega _{j}^{2}\le j(N-j)$ for all $j$ from 
theorem \ref{mtheorem}, it follows that 
$\omega _{j}^{2}(\tau )=j(N-j)$ for sufficiently large $\tau $
and all $j$.  
Therefore $U_{j}(\tau )$ is zero for sufficiently large $\tau $,
hence $m(\tau )$ is bounded as $\tau \rightarrow \infty $ and
$\lim _{\tau \rightarrow \infty } \Psi (\tau )=1$.

\bigskip
The remaining intermediate case is where $\omega _{j}$ and
$\omega _{j+l}$, where $l>1$, are monotonic for all 
sufficiently large $\tau $,
whilst $\omega _{j+i}^{2}$, $i=1,\ldots ,l-1$ have minima for
arbitrarily large $\tau $.
We include the possibility that $j=0$,  to cover the case where
there is only one $\omega _{i}$ which has a limit.
Define $L_{1}$ and $L_{2}$ by
\be
\omega _{j}^{2} \rightarrow L_{1}, \qquad
\omega _{j+l}^{2} \rightarrow L_{2} 
\ee
as $\tau \rightarrow \infty $.
Given $\epsilon >0$, there is a $T_{1}$ such that
\be
\left| \omega _{j}^{2}-L_{1} \right| <\epsilon ,
\qquad
\left| \omega _{j+l}^{2}-L_{2} \right| <\epsilon
\qquad 
\forall \tau > T_{1}.
\ee   
If $\omega _{j+1}^{2}$ has a minimum at $\tau _{0}>T_{1}$, then
\bea
\omega _{j+1}^{2}(\tau _{0}) & \ge & 
1+\frac {1}{2} \left( \omega _{j+2}^{2}(\tau _{0}) +
\omega _{j}^{2}(\tau _{0}) \right) 
\nonumber
\\
 & \ge & 1+\frac {1}{2} \left(
\omega _{j+2}^{2}(\tau _{0}) +L_{1} -\epsilon \right) .
\eea
With $N_{j+1}$ as before, then
\be
N_{j+1} \ge 1+\frac {1}{2} N_{j+2} +K_{1}
\ee
where $K_{1}=\frac {1}{2} \left( L_{1}-\epsilon \right)$.
Similarly,
\be
N_{j+l-1}\ge 1+\frac {1}{2} N_{j+l-2}+K_{2}
\ee
where $K_{2}=\frac {1}{2} \left( L_{2}-\epsilon \right) $.
The remaining inequalities for $N_{j+2},\ldots ,N_{j+l-2}$
are exactly as before (\ref{nineq}).
The previous method can be repeated to give
\be
N_{j+i}\ge i(l-i)+K_{1}+K_{2} \qquad i=1,\ldots ,l-1.
\ee

\bigskip
Now define ${\tilde {N}}_{i}$ to be the lowest upper bound of the 
set of maximum values of $\omega _{i}^{2}$ (which exists since
$\omega _{i}^{2}$ is bounded).
If $\omega _{j+1}^{2}$ has a maximum at $\tau _{0}>T_{1}$, then
\bea
\omega _{j+1}^{2}(\tau _{0}) & \ge & 
1+\frac {1}{2} \left( \omega _{j+2}^{2}(\tau _{0})+
\omega _{j}^{2}(\tau _{0}) \right)
\nonumber
\\
& \ge & 1+\frac {1}{2} \left(
\omega _{j+2}^{2} +L_{1}+\epsilon  \right) .
\eea
The analysis now proceeds exactly as in the previous paragraph
and yields
\be
{\tilde {N}}_{j+i}\le i(l-i)+{\tilde {K}}_{1}+{\tilde {K}}_{2}
\qquad
i=1,\ldots ,l-1
\ee
where ${\tilde {K}}_{i}=\frac {1}{2} \left( L_{i}+\epsilon \right)$ for
$i=1,2$.
In addition, by definition it must be the case that
\be
{\tilde {N}}_{j+i}\ge N_{j+i}
\ee
which means that
\be
i(l-i)+\frac {1}{2} \left( L_{1}+L_{2} \right) -\epsilon
\le N_{j+i} \le {\tilde {N}}_{j+i} \le
i(l-i)+\frac {1}{2} \left( L_{1}+L_{2} \right) +\epsilon
\ee
for all $\epsilon >0$, whence ${\tilde {N}}_{j+i}=N_{j+i}$ for
all $i$ and
\be
\omega _{j+i}^{2} =i(l-i)+\frac {1}{2} \left( L_{1}+L_{2} \right)
\ee
for sufficiently large $\tau $.

\bigskip
In conclusion, then, we have shown that the intermediate case has
some $\omega _{j}$'s which are monotonic for sufficiently 
large $\tau $, with the remaining $\omega _{j}$'s being
constant for sufficiently large $\tau $. 
Hence in this case also $m$ is bounded as $\tau \rightarrow \infty $
and $\lim _{\tau \rightarrow \infty }\Psi (\tau )=1$.
\hfill
$\square $

\bigskip
\noindent
{\bf {Proof of Proposition \ref{flatprop}}}
\smallskip
\newline
In order to show that the geometry becomes flat as 
$\tau \rightarrow \infty $, it remains to show that $S\rightarrow 1$
as $\tau \rightarrow \infty $.
From the proof of lemma \ref{flatlem}, for $\omega _{j}$
having a limit as $\tau \rightarrow \infty $, we have
\be
rU_{i}={\dot {\omega }}_{i}\rightarrow 0 
\ee
as $\tau \rightarrow \infty $.
In case 1 of the proof of lemma \ref{flatlem}, the proof
of \cite{bfm} carries straight over to show that 
$U_{i}\rightarrow 0$ as $\tau \rightarrow \infty $ also in this
situation.
Now
\be
\frac {{\dot {S}}}{S}=\Psi G = 
\frac {1}{\Psi } \sum _{i=1}^{N-1} U_{i}^{2} \le
\frac {C}{\Psi r^{2}}
\ee
for some constant $C$ and sufficiently large $\tau $.
Therefore
\be
\frac {S'}{S}=\frac {{\dot {S}}}{S} \frac {1}{r\Psi } \le
\frac {C}{\mu r^{3}}
\le \frac {{\tilde {C}}}{r^{3}}
\ee
for some constant ${\tilde {C}}$ and sufficiently large $\tau $,
as $\mu \rightarrow 1$ as $\tau \rightarrow \infty $.
This means that $S$ has a finite limit as $\tau \rightarrow \infty $.

\bigskip
The field equations only involve $\frac {S'}{S}$ rather than just $S$
itself. 
Therefore $S$ is defined only up to a multiplicative constant,
and without loss of generality we may therefore take the
finite limit of $S$ as $\tau \rightarrow \infty $ to be 1,
so that the spacetime is asymptotically flat.

\bigskip
Let $\delta (\tau )=2\Psi - \kappa -1$ be a small perturbation,
then the equation for $\omega _{j}$ reads
\be
{\ddot {\omega }}_{j}+\left[ 1 -\omega _{j}^{2} 
+\frac {1}{2} \left( \omega _{j+1}^{2} +\omega _{j-1}^{2} \right)
\right] \omega _{j} = \left( 1+ \delta (\tau ) \right)
{\dot {\omega }}_{j}
\ee
as $\tau \rightarrow \infty $.
Since $\delta $ is small, it does not alter the position or nature
of the critical points as compared with the exactly 
flat space case. 
Lemma \ref{flatlem} showed that each  
${\dot {\omega }}_{j},{\ddot {\omega }}_{j}\rightarrow 0$ 
as $\tau \rightarrow \infty $.
Hence $\omega _{j}$ must approach one of the critical points 
whose nature was elucidated in section \ref{flat}.
The flat space analysis showed that, if $\omega _{j}=0$,
then there is an unstable focal point in the 
$(\omega _{j},{\dot {\omega }}_{j})$ plane.
Hence our solution cannot approach this value, unless 
$\omega _{j}\equiv 0$.
The solution has to tend to one of the saddle points along a
stable direction, and, in direct analogy with the 
${\mathfrak {su}}(2)$ case \cite{ershov}, 
there are no solutions where $\omega _{j}\rightarrow 0$
as $\tau \rightarrow \infty $.
The solution must therefore be a member of the family found in the
local existence theorem \ref{infprop}.
\hfill
$\square $

\bigskip
The power of proposition \ref{flatprop} lies in that, if we can prove
the existence of solutions for which $\Psi >0$ for all $\tau >0$
and $r\rightarrow \infty $, then these solutions are automatically 
the regular black holes (and, if $r_{h}\rightarrow 0$, 
soliton solutions \cite{bartnik})
we seek. 
We close this section by determining the asymptotic behaviour of
solutions of type 2.

\begin{prop}
\label{rmaxprop}
If $\Psi (\tau )>0$ for all $\tau >0$ and $r(\tau )$ remains 
bounded, then $\Psi \rightarrow 0$ and $\kappa \rightarrow 1$
as $\tau \rightarrow \infty $.
In addition, there is at least one $j$ for which
$\omega _{j} \rightarrow 0$ as
$\tau \rightarrow \infty $, and this $\omega _{j}$ has
infinitely many zeros.
\end{prop}

\proof
Since $r$ is monotonic increasing and bounded, it has a limit $r_{0}$.
In addition, $m$ is monotonic increasing and bounded since
the positivity of $\mu $ implies that
\be
m<\frac {r}{2} \le \frac {r_{0}}{2}.
\ee
As $\tau \rightarrow \infty $, ${\dot {r}}\rightarrow 0$
and hence $\Psi \rightarrow 0$ from (\ref{fregeqn}).
Consider the quantity
\be
E=-\frac {r^{2}}{4}\left( 1+\Psi ^{2} -2\kappa \Psi \right)
=-\frac {r^{2}}{2} \left( -\mu G+r^{2} p_{\theta } \right).
\ee
Then
\bea
{\dot {E}} & = & 2r^{2}\mu G \Psi -r^{2}\kappa \mu G
\nonumber 
\\
& < &
 r^{2}\mu G (1-\kappa) 
\nonumber
\\
& < & 0,
\eea
for sufficiently large $\tau $ for which $\Psi < 1/2$, and since
$\kappa >1$ \cite[Lemma 10]{bfm}.
Hence 
\be
E\rightarrow -\frac {r_{0}^{2}}{4}
\label{elim}
\ee
as $\tau \rightarrow \infty $, since $E$ is monotonically decreasing
for sufficiently large $\tau $.
Then, following \cite[Proposition 15]{bfm}, we have 
$\kappa \rightarrow 1$ as $\tau \rightarrow \infty $.

\bigskip
Let $\delta (\tau )=\kappa -2\Psi -1$ be small, then the
equation for $\omega _{j}$ reads
\be
{\ddot {\omega }}_{j} = {\dot {\omega }}_{j} 
\left( -1-\delta \right) - \omega _{j} \left[
1-\omega _{j}^{2} +\frac {1}{2} \left(
\omega _{j+1}^{2} +\omega _{j-1}^{2} \right) \right]  .
\ee
Replacing $\tau $ by $-\tau $ yields the equation
\be
{\ddot {\omega }}_{j} = {\dot {\omega }}_{j} 
\left( 1+\delta \right) - \omega _{j} \left[
1-\omega _{j}^{2} +\frac {1}{2} \left(
\omega _{j+1}^{2} +\omega _{j-1}^{2} \right) \right]  .
\ee
which is the equation for flat space solutions, so that $\omega _{j}$
must approach one of the critical points.
The fact that $E\neq 0$ means that $p_{\theta }$ cannot tend to zero
as $\tau \rightarrow \infty $ because $\mu G$ does vanish in the limit.
Hence at least one of the $\omega _{j}$'s must be zero as
$\tau \rightarrow \infty $.
Since $\delta $ is small it alters neither the position nor
the characteristics of the critical points, and hence this 
$\omega _{j}$ will go into the focus at zero. 
Therefore it oscillates infinitely many times
 before it hits zero.
\hfill
$\square $

\bigskip
We may determine the value of $r_{0}$ as follows.
Since $\kappa $ is finite and we have (\ref{elim}), it follows
that $r^{2}p_{\theta }\rightarrow 1/2$ as
$\tau \rightarrow \infty $.
Hence
\be
r_{0}^{2}=\frac {1}{2} \lim _{\tau \rightarrow \infty }
\sum _{j=1}^{N}
\left( \omega _{j}^{2}-\omega _{j-1}^{2} -N-1+2j \right) ^{2}.
\ee
The maximum possible value of $r_{0}^{2}$ is when all
$\omega _{j}\rightarrow 0$ as $\tau \rightarrow \infty $.
In this case:
\be
r_{0,max}^{2} =\frac {1}{2} \sum _{j=1}^{N} \left(
-N-1+2j \right) ^{2} = \frac {1}{6} N(N-1)(N+1).
\ee
The minimum possible value of $r_{0}^{2}$ is when only one
$\omega _{j}\rightarrow 0$ as $\tau \rightarrow \infty $, when
\be
r_{0}^{2} =j^{2}(N-j)^{2}
\ee
which has a minimum when $j=1$ or $N-1$, and
\be
r_{0}^{2} =(N-1)^{2}.
\ee
With this value of $r_{0}$, we follow \cite{bfm} and denote
by $S_{\infty }$ singular solutions for which $\mu $ vanishes
outside the event horizon when $r<r_{0,max}$.

\section{${\mathfrak {su}}(3)$ black holes}
\label{su3}
We are now in a position to prove the existence of regular black
holes.
We begin in this section with ${\mathfrak {su}}(3)$, 
which we shall study in some
detail, before proceeding to the general case in the next section.
The method in general is exactly the same as in this section, 
but it is hoped that by doing the simplest specific case,
where we have just two gauge field degrees of freedom, 
explicitly, the proof will be more transparent.

\bigskip
Genuine ${\mathfrak {su}}(3)$ solutions (as opposed to 
embedded ${\mathfrak {su}}(2)$ solutions which are discussed below)
have been found numerically in \cite{kunz1}.
There it was conjectured that, analogous to the ${\mathfrak {su}}(2)$
case, black hole solutions exist for an infinite, but discrete,
set of points in the parameter space given by the values 
at the event horizon of the
functions describing the gauge degrees of freedom.

\bigskip
Firstly, we review briefly the way in which 
${\mathfrak {su}}(2)$ solutions may
be embedded in ${\mathfrak {su}}(3)$.  
For neutral black holes, numerical solutions are discussed in 
\cite{klei2}, where both ${\mathfrak {su}}(2)$ and 
${\mathfrak {so}}(3)$ embeddings are discussed.
The ${\mathfrak {so}}(3)$ embedding contains scaled 
${\mathfrak {su}}(2)$ black holes as well as genuine 
${\mathfrak {so}}(3)$ solutions possessing two degrees of 
freedom, which are indexed by the number of nodes of each of 
the gauge field functions.
The method for embedding charged black holes can be
found in \cite{klei1}, where the various possible embeddings 
are described in detail and numerical solutions 
presented for the various embeddings in ${\mathfrak {su}}(5)$.
Such solutions possess Coulomb charge arising from one or more
of the gauge degrees of freedom. 
The remaining gauge field functions do not contribute to the
charge and have structures similar to the neutral solutions.
The black holes are once again indexed by the number of nodes of
the gauge functions.

\bigskip
\noindent
{\em {Neutral black holes}}

\smallskip
\noindent
Let $\omega _{1}(r)={\sqrt {2}}\omega (r)$ and
$\omega _{2}(r)={\sqrt {2}}\omega (r)$.
Then the field equations (\ref{firsteqn}--\ref{lasteqn}) become:
\bea
r^{2} \mu \omega '' & = & 
-\left( 2m -2r^{3} p_{\theta } \right) \omega '
-\left( 1-\omega ^{2} \right) \omega ,
\\
m' & = & 
\left( \mu G +r^{2} p_{\theta } \right) ,
\\
\frac {S'}{S} & = & \frac {2G}{r},
\eea
where
\be
\mu = 1-\frac {2m}{r}, \qquad
G=4\omega ^{'2}, \qquad
p_{\theta }= \frac {2}{r^{4}} \left( \omega ^{2}-1 \right) ^{2}.
\ee
In order to extract the precise form of the 
${\mathfrak {su}}(2)$ equations, define a new independent 
variable $R$ and a new dependent variable $M$ by
\be
R=\frac {1}{2} r, \qquad
M=\frac {1}{2} m,
\ee
with the other field variables remaining unchanged.
Then the equations are
\bea
R^{2} \mu \frac {d^{2}\omega }{dR^{2}} & = & 
-\left[ 2M -\frac {(\omega ^{2}-1)^{2}}{2R} \right] 
\frac {d\omega }{dR} - \left( 1-\omega ^{2} \right) \omega ,
\\
\frac {dM}{dR} & = & \left[
\mu \left( \frac {d\omega }{dR} \right) ^{2} +
\frac {1}{2R^{2}} \left( \omega ^{2}-1 \right) ^{2} \right]
\\
\frac {1}{S} \frac {dS}{dR}  & = &
\frac {2}{R} \left( \frac {d\omega }{dR} \right) ^{2}.
\eea
These are now exactly the ${\mathfrak {su}}(2)$ field equations,
and the existence of regular black hole solutions has been 
proved in \cite{bfm,sw}.
 
\bigskip
\noindent
{\em {Charged black holes}}

\smallskip
\noindent
Suppose $\omega _{2}(r)\equiv 0$ (the case 
$\omega _{1}(r) \equiv 0$ is identical).
Then
\bea
r^{2} \mu \omega _{1}'' & = &
-\left( 2m -2r^{3} p_{\theta } \right) \omega _{1}'
-\left( 1-\omega _{1}^{2} \right) \omega _{1},
\\
m' & = & 
\left[
\omega _{1}^{'2} +\frac {1}{2r^{2}} \left(
\omega _{1}^{2} -1 \right) ^{2} +\frac {3}{2r^{2}} \right] ,
\\
\frac {S'}{S} & = & \frac {2\omega _{1}^{'2}}{r}.
\eea
While these equations are {\em {not}} the same as those for
${\mathfrak {su}}(2)$ neutral black holes, the equations
for $\kappa $, $\Psi $, $U$, $\omega_{1} $, $r$ and $S$
as functions of $\tau $ {\em {are}} identical.
Hence the analysis of \cite{bfm} carries over immediately to 
prove the existence of black hole solutions of this form.

\bigskip
Our first existence proposition states the existence of
genuine ${\mathfrak {su}}(3)$ black holes, that is, black holes
which do not belong to one of the two families outlined above.
The proof is remarkably simple once we have set up our notation
suitably.

\begin{prop}
\label{firstprop}
There exist regular black hole solutions of the ${\mathfrak {su}}(3)$
Einstein-Yang-Mills equations which do not correspond to 
embedded ${\mathfrak {su}}(2)$ charged or neutral black holes.
\end{prop}

The strategy in proving this proposition is very similar to that
used in \cite{bfm}.
Generic values of the starting parameters 
$(\omega _{1,h},\omega _{2,h})$ lead to a singular solution,
but we will prove that there must be some values of the starting
parameters which do not lead to a singular solution.  
Firstly, we rule out the possibility of solutions of type 2.
From section \ref{global}, any solution for which $\mu (\tau )>0$
for all $\tau $ and $r(\tau )$ remains bounded as 
$\tau \rightarrow \infty $, has $r\rightarrow r_{0}$ as
$\tau \rightarrow \infty $, where
\be
(N-1)^{2} \le r_{0}^{2} \le 
\frac {1}{6} N(N+1)(N-1).
\ee
For $N=3$, therefore, $r_{0}=2$. 
Fixing $r_{h}>2$ will therefore rule out the possibility of such
solutions.
This is analogous to the value $r_{h}>1$ which rules out
such solutions in the ${\mathfrak {su}}(2)$ case.
At the end of this section we shall return to the existence proof
for $r_{h}\le 2$.

\bigskip
For every pair of starting values $(\omega _{1,h},\omega _{2,h})$
there are then just two possibilities:
\begin{enumerate}
\item
$\mu (\tau )>0$ for all $\tau $ and we have a regular black hole
solution;
\item
there is a $\tau _{0}>0$ such that $\mu (\tau _{0})=0$
and we have a singular solution.
\end{enumerate}
Define a new variable $R$ by
\be
R=\frac {r-r_{h}}{r}
\ee
and $R_{m}$ by the maximum value of $R$ for each solution, that is,
in case one $R_{m}=1$ (corresponding to $r\rightarrow \infty $), and
in case two $R_{m}=R(\tau _{0})$.
For each solution we define $n_{i}$ to be the number of zeros of
the function $\omega _{i}$ between $\tau =0$ and $\tau =\infty $
in case 1 and $\tau =\tau _{0}$ in case 2, where $n_{i}$ can be
infinite.
Defining new variables $N_{i}$ by
\be
N_{i}=\frac {n_{i}}{1+n_{i}}
\ee
we allow the possibility that $N_{i}=1$.

\bigskip
Each pair of non-zero starting values $(\omega _{1,h},\omega _{2,h})$
can then be mapped to the three quantities
$(R_{m},N_{1},N_{2})$ for the corresponding solution, giving
rise to a map from ${\mathbb {R}}^{2}$ 
($(\omega _{1,h},\omega _{2,h})$ space) to 
${\mathbb {B}}^{2}\times [0,1]$ ($(R_{m},N_{1},N_{2})$ space),
where ${\mathbb {B}}$ is the discrete set
$\{ 0, 1/2, 2/3, 3/4, \ldots , 1\} $.
Call this map $f$.
Figures 1--3 sketch the nature of $(\omega _{1,h},\omega _{2,h})$
space, ${\mathbb {B}}\times [0,1]$ and ${\mathbb {B}}^{2}$
respectively.
Note that this map will not be 1-1 for $N=3$, although
it was conjectured in \cite{bfm} that for $N=2$ the map is 1-1.

\bigskip
Note that since we know the nature of the solution space 
when one of the $\omega _{i,h}$ vanishes (since in this case 
$\omega _{i}\equiv 0$), we do not need to extend the map $f$
to the co-ordinate axes in $(\omega _{1,h},\omega _{2,h})$
space.
With this notation, we are ready to prove proposition
\ref{firstprop}, once we have the following lemma.

\begin{lemma}
\label{fcont}
Suppose that for starting values
$({\bar {\omega }}_{1,h},{\bar {\omega }}_{2,h})$
there is a singular solution with $\mu ({\bar {\tau }}_{0})=0$
and the gauge field functions having node structure 
$({\bar {n}}_{1},{\bar {n}}_{2})$.
Then the map $f$ is continuous at 
$({\bar {\omega }}_{1,h},{\bar {\omega }}_{2,h})$.
\end{lemma}

\proof
From proposition \ref{horprop}, the field variables are 
continuous
in $\tau $ and the starting parameters.
Thus all starting values 
$(\omega _{1,h},\omega _{2,h})$ in a sufficiently small 
neighbourhood of 
$({\bar {\omega }}_{1,h},{\bar {\omega }}_{2,h})$
will give rise to a singular solution with $\mu (\tau _{0})=0$
where $\tau _{0}$ is close to ${\bar {\tau }}_{0}$,
the values of the field variables at $\tau _{0}$ will be
close to those at ${\bar {\tau }}_{0}$ in the original solution
and the node structure will be $({\bar {n}}_{1},{\bar {n}}_{2})$
since the gauge field functions cannot have double zeros
(proposition \ref{maxminprop}).
In other words, $f$ is continuous at
$({\bar {\omega }}_{1,h},{\bar {\omega }}_{2,h})$.
\hfill
$\square $

\bigskip
\noindent
{\bf {Proof of Proposition \ref{firstprop}}}
\smallskip
\newline
Consider the open subset of the $(\omega _{1,h},\omega _{2,h})$
plane given by
\be
D=\{ (\omega _{1,h},\omega _{2,h}) : 
0<\omega _{1,h}<\omega _{2,h} \} .
\ee
The subset $D'=\{ (\omega _{1,h},\omega _{2,h}):
0<\omega _{2,h}<\omega _{1,h} \} $
can be treated similarly.
The symmetries of the field equations (\ref{firsteqn}--\ref{lasteqn})
mean that it is sufficient to consider only the first
quadrant of the $(\omega _{1,h},\omega _{2,h})$ plane.
From \cite{bfm} we know that along the line
$\omega _{1,h}=\omega _{2,h}$ there are singular solutions with
node structure $(n_{1},n_{1})$ for all $n_{1}=1,2,\ldots $.
Therefore there are neighbourhoods of $D$ corresponding
to singular solutions with node structure $(n_{1}, n_{1})$
for all $n_{1}=1,2,\ldots $.
Hence $f(D)$ contains points corresponding to $(R_{m},N_{1},N_{1})$
for all $N_{1}\in {\mathbb {B}}$.
Therefore $f(D)$ is not a connected set.
Therefore $f$ cannot be continuous everywhere on $D$ because $D$ is
connected.
Hence we conclude that there exists at least one 
$(\omega _{1,h},\omega _{2,h})\in D$ corresponding to a black hole
solution because $f$ is continuous for all 
$(\omega _{1,h},\omega _{2,h})$ corresponding to singular solutions.
\hfill
$\square $

\bigskip
This approach allows us to see quite simply how the transversality
property conjectured in \cite{bfm} arises in the ${\mathfrak {su}}(2)$
case.
Along the line of values of $\omega _{h}$, singular solutions
having different numbers of nodes must be separated by at least
one regular solution.  
For ${\mathfrak {su}}(2)$, it is known that values of
$\omega _{h}$ sufficiently close to $1$ correspond to singular
solutions with $\omega $ having one node,
whilst sufficiently small $\omega _{h}$ correspond to solutions
having as many nodes as we like.
In addition, if we have a regular solution with $n$ nodes,
then all singular solutions with $\omega _{h}$
sufficiently close have either $n$ or $n+1$ nodes.
Starting with the singular solutions with one node, and decreasing
$\omega _{h}$, we first hit a regular solution with one node
(there being only the trivial solution with no nodes).
Then there may be more singular solutions with one node, or two nodes.
In the former case there must be another regular solution with
one node, in the latter case there will next be a regular solution
with two nodes.
Either way, there must be a regular solution with two nodes before
we can move on to singular solutions with $n>2$ nodes.
The process continues for all $n$.

\bigskip
Having shown that there exist genuine ${\mathfrak {su}}(3)$
black holes, the main result of this section is the following
theorem, which we prove at the  moment for the case $r_{h}>2$,
returning to the case $r_{h}\le 2$ at the end of the section.

\begin{theorem}
\label{exist}
Given ${\bar {n}}_{1}=0,1,\ldots $, then there exist
regular black hole solutions of the ${\mathfrak {su}}(3)$
Einstein-Yang-Mills equations with $\omega _{1}(r)$ having
${\bar {n}}_{1}$ nodes and $\omega _{2}(r)$ having $n_{2}$ nodes,
for infinitely many $n_{2}\ge {\bar {n}}_{1}$.
\end{theorem}

A similar result holds with the roles of $\omega _{1}$ and
$\omega _{2}$ reversed. 
Note that this result is slightly weaker than the corresponding
theorem for ${\mathfrak {su}}(2)$, since we cannot guarantee that
every combination of $(n_{1},n_{2})$ is the node structure of
the gauge fields for some black hole solution, only that an infinite
number of such combinations does occur for each $n_{1}$.
At the end of this section we shall give an argument, based on
a numerical analysis, that in fact black holes exist for all
$(n_{1},n_{2})$, although we are not able to prove this
analytically.
The proof of theorem \ref{exist} will proceed via a series of lemmas.

\begin{lemma}
\label{weaklem}
Suppose that the starting parameters $(0,{\bar {\omega }}_{2,h})$
correspond to a charged regular black hole solution in which
$\omega _{2}(r)$ has ${\bar {n}}_{2}$ zeros.
Then, given $n_{0}$, for all sufficiently small $\omega _{1,h}$
and $\omega _{2,h}$ sufficiently close to ${\bar {\omega }}_{2,h}$,
the solutions are regular or singular solutions with 
$n_{1}\ge n_{0}$ and $n_{2}\ge {\bar {n}}_{2}$.
\end{lemma}

\proof
By continuity, all field variables will remain close to the original
charged solution until $r\gg 1$ and the geometry is approximately
flat.
At this point $\omega _{1}$ will be very small and $\omega _{2}$
will be close to $1$.
The equations for $\epsilon _{1}=\omega _{1}$ and 
$\epsilon _{2}=\omega _{2}-1$ as functions of $\tau $ are, in 
this regime, to first order,
\bea
0 & = & {\ddot {\epsilon }}_{1}-{\dot {\epsilon }}_{1}
+\frac {3}{2}\epsilon _{1} 
\nonumber
\\
0 & = & {\ddot {\epsilon }}_{2} -{\dot {\epsilon }}_{2}
-2\epsilon _{2}.
\eea
This corresponds to a focus in the 
$(\omega _{1},{\dot {\omega }}_{1})$ plane, and hence with 
$\omega _{1,h}$ sufficiently small, $\epsilon _{1}$ will have
at least $n_{0}$ zeros.
\hfill
$\square $

\begin{lemma}
\label{reglem}
If $({\bar {\omega }}_{1,h},{\bar {\omega }}_{2,h})$
leads to a regular black hole solution with $\omega _{1}$ having
${\bar {n}}_{1}$ nodes and $\omega _{2}$ having ${\bar {n}}_{2}$
nodes, then all $(\omega _{1,h},\omega _{2,h})$ sufficiently 
close to
$({\bar {\omega }}_{1,h},{\bar {\omega }}_{2,h})$
lead to regular or singular solutions with $\omega _{1}$ having
at least ${\bar {n}}_{1}$ zeros and $\omega _{2}$ having
at least ${\bar {n}}_{2}$ zeros.
\end{lemma}

\proof
Since the solutions are continuous in the starting parameters and
$r$, for $(\omega _{1,h},\omega _{2,h})$ sufficiently close to
$({\bar {\omega }}_{1,h},{\bar {\omega }}_{2,h})$,
the gauge function $\omega _{1}$ will have ${\bar {n}}_{1}$
zeros and $\omega _{2}$ will have ${\bar {n}}_{2}$ zeros for 
$r<r_{1}$ for some $r_{1}$ and $\omega _{i}(r_{1})$ will be
close to their asymptotic values.
\hfill
$\square $
 
\bigskip
From the proof of proposition \ref{firstprop}, 
it is expected that the 
$(\omega _{1,h},\omega _{2,h})$ plane will be partitioned into 
open sets containing singular solutions with $(n_{1},n_{2})$
nodes by lines of regular solutions.
These lines of regular solutions will occur as the horizon
radius, $r_{h}$ is varied.  
Numerical investigations \cite{klei1,klei2,kunz1}
indicate that for fixed $r_{h}$ regular solutions exist at discrete
points in the $(\omega _{1,h},\omega _{2,h})$ plane, the positions
of these discrete points varying as $r_{h}$ is varied.
Lemma \ref{reglem} shows that lines corresponding to 
regular solutions with different numbers of nodes must be
disjoint, and furthermore that the region corresponding to
singular solutions with $({\bar {n}}_{1},{\bar {n}}_{2})$ nodes
will be bounded by lines of regular solutions having
$(n_{1},n_{2})$ nodes, where $n_{1}\le {\bar {n}}_{1}$
and $n_{2}\le {\bar {n}}_{2}$.
Lemma \ref{reglem} is somewhat weaker than the corresponding
result for ${\mathfrak {su}}(2)$ \cite[Proposition 22]{bfm},
since the coupling between the $\omega $'s means that we
cannot bound from above the number of zeros of the gauge field
functions by analytic arguments.
Below we shall address this question further by numerical
investigations. 
This leads directly to our existence theorem \ref{exist}
being slightly weaker than for $N=2$.


\begin{lemma}
\label{singlem}
Suppose $(0,{\bar {\omega }}_{2,h})$ corresponds to a charged regular
solution, where $\omega _{2}$ has ${\bar {n}}_{2}$ nodes.
Given $n_{0}$, let $D_{0}$ be the largest neighbourhood of
$(0,{\bar {\omega }}_{2,h})$ for which $(\omega _{1,h},\omega _{2,h})$
correspond to regular or singular solutions with $n_{1}\ge n_{0}$
and $n_{2}\ge {\bar {n}}_{2}$.
Suppose, in addition, that there is 
$({\hat {\omega }}_{1,h},{\hat {\omega }}_{2,h})\in D_{0}$
corresponding to a singular solution with $n_{1}=n_{0}$
and $n_{2}={\bar {n}}_{2}$.
Then there is $(\omega _{1,h}^{R},\omega _{2,h}^{R})\in D_{0}$
corresponding to a regular solution with $n_{1}=n_{0}$
and $n_{2}={\bar {n}}_{2}$.
\end{lemma}

\proof
Let ${\tilde {n}}_{0}>n_{0}$.
Then there is a neighbourhood 
${\tilde {D}}_{0}\subset D_{0}, {\tilde {D}}_{0}\neq D_{0}$
for which all solutions are regular or singular solutions with
$n_{1}\ge {\tilde {n}}_{0}$ and $n_{2}\ge {\bar {n}}_{2}$.
In ${\tilde {D}}_{0}$ there must be at least one singular solution
having $n_{1}\ge {\tilde {n}}_{0}$ and $n_{2}\ge {\bar {n}}_{2}$,
by lemma \ref{reglem}, corresponding to starting values
$({\tilde {\omega }}_{1,h},{\tilde {\omega }}_{2,h})$.
Consider a curve joining 
$({\tilde {\omega }}_{1,h},{\tilde {\omega }}_{2,h})$
and $({\hat {\omega }}_{1,h},{\hat {\omega }}_{2,h})$
and lying in $D_{0}$.
Then there must be at least one regular solution along this curve.
Let $(\omega _{1,h}^{R},\omega _{2,h}^{R})$
be the regular solution closest to 
$({\hat {\omega }}_{1,h},{\hat {\omega }}_{2,h})$,
having $n_{1}=n_{1}^{R}$ and $n_{2}=n_{2}^{R}$.
Since $(\omega _{1,h}^{R},\omega _{2,h}^{R})\in D_{0}$, it follows 
that $n_{1}^{R}\ge n_{0}$ and $n_{2}^{R}\ge {\bar {n}}_{2}$.
Also, from lemma \ref{reglem}, 
$n_{1}^{R}\le n_{0}$ and $n_{2}^{R}\le {\bar {n}}_{2}$,
since sufficiently close to 
$(\omega _{1,h}^{R},\omega _{2,h}^{R})$ there are singular solutions
having $n_{1}=n_{0}$ and $n_{2}={\bar {n}}_{2}$.
Therefore $n_{1}^{R}=n_{0}$ and $n_{2}^{R}={\bar {n}}_{2}$.
\hfill
$\square $

\begin{lemma}
\label{existlem}
Given ${\bar {n}}_{2}$ and $n_{0}>{\bar {n}}_{2}$, then there exist
regular and singular solutions with $n_{2}={\bar {n}}_{2}$ and
$n_{1}\ge n_{0}$.
\end{lemma}

\proof
Consider the point $(0,{\bar {\omega }}_{2,h})$ which corresponds
to the charged regular solution with $n_{2}={\bar {n}}_{2}$.
Then all $(\omega _{1,h},\omega _{2,h})$ sufficiently close to
$(0,{\bar {\omega }}_{2,h})$ are regular or singular solutions
with $n_{1}\ge n_{0}$ and $n_{2}\ge {\bar {n}}_{2}$.
Let all such $(\omega _{1,h},\omega _{2,h})$ form a neighbourhood
$D_{0}$ of $(0,{\bar {\omega }}_{2,h})$.
From \cite{bfm}, there exist $(0,{\hat {\omega }}_{2,h})\in D_{0}$
corresponding to singular solutions having $\omega _{1}\equiv 0$
and $n_{2}={\bar {n}}_{2}$.
By continuity, all $(\omega _{1,h},\omega _{2,h})$ sufficiently close to
$(0,{\hat {\omega }}_{2,h})$ correspond to singular solutions having
$n_{1}\ge n_{0}$ and $n_{2}={\bar {n}}_{2}$.
Hence, by lemma \ref{singlem}, there are also in this neighbourhood
regular black hole solutions having $n_{1}\ge n_{0}$ and
$n_{2}={\bar {n}}_{2}$.
\hfill
$\square $

\bigskip
\noindent 
{\bf {Proof of Theorem \ref{exist}}} 
\smallskip 
\newline 
Fix ${\bar {n}}_{2}$.
Then we have regular solutions with node structure 
$({\bar {n}}_{2}, {\bar {n}}_{2})$.
Now let $n_{0}={\bar {n}}_{2}+1$, then from lemma \ref{existlem}
there are regular solutions having $n_{2}={\bar {n}}_{2}$ and
$n_{1}\ge n_{0}$.
Let $n_{0}'$ be the smallest such $n_{1}$.
Now set $n_{0}=n_{0}'+1$ and repeat the process.
\hfill
$\square $

\bigskip
In order to guarantee the existence of black hole solutions having
node structure $({\bar {n}}_{1},{\bar {n}}_{2})$ for every pair of
integers $({\bar {n}}_{1},{\bar {n}}_{2})$, we would require the
following lemma in addition to lemma \ref{existlem}.

\begin{lemma}
\label{tightlem}
Suppose $({\bar {\omega }}_{1,h},{\bar {\omega }}_{2,h})$
corresponds to a regular black hole solution in which
$\omega _{1}$ has ${\bar {n}}_{1}$ nodes and
$\omega _{2}$ has ${\bar {n}}_{2}$ nodes.
Then, for $(\omega _{1,h},\omega _{2,h})$ sufficiently close
to $({\bar {\omega }}_{1,h},{\bar {\omega }}_{2,h})$,
solutions for which $\omega _{2}$ still has ${\bar {n}}_{2}$
nodes are such that $\omega _{1}$ has either ${\bar {n}}_{1}$
or ${\bar {n}}_{1}+1$ nodes.
\end{lemma}

The argument we now give for lemma \ref{tightlem} is not an analytic
proof because we require a numerical analysis.

\bigskip
\proof
By continuity, we know that for $(\omega _{1,h},\omega _{2,h})$
sufficiently close to 
$({\bar {\omega }}_{1,h},{\bar {\omega }}_{2,h})$,
the gauge function $\omega _{1}$ has ${\bar {n}}_{1}$ zeros
and $\omega _{2}$ has ${\bar {n}}_{2}$ zeros for $r<r_{1}$ for
some $r_{1}$ and $\omega _{i}(r_{1})$ will be close to their
asymptotic values.
Since we are considering only solutions for which $\omega _{2}$
has ${\bar {n}}_{2}$ nodes, for $r>r_{1}$, then, $\omega _{2}$ will
be of one sign.
Therefore we may consider the new dependent variable
\be
\psi = \frac {\omega _{1}}{\omega _{2}}
\ee
which will have the same number of zeros as $\omega _{1} (r)$
for $r>r_{1}$.
Then the equation satisfied by $\psi $ is
\be
r^{2} \mu \psi '' +\left( 2m-2r^{3}p_{\theta } \right)
\psi ' +2r^{2}\mu \frac {\omega _{2}'}{\omega _{2}} 
\psi '
+\frac {3}{2} \omega _{2}^{2} \psi \left( 1-\psi ^{2} \right) =0.
\label{psieqn}
\ee
On some interval, $r_{1}<r<r_{2}$, the geometry will be very nearly
flat, and it will remain flat until one of $\omega _{1}$, 
$\omega _{2}$ or their derivatives approach $O({\sqrt {r}})$.
Next consider the flat space equations for $\psi $ and $\omega _{2}$,
which have the autonomous form (where $\tau = \log r$):
\bea
0 & = & 
{\ddot {\omega }}_{2}-{\dot {\omega }}_{2} +\left( 1-\omega _{2}^{2}
+\frac {1}{2} \psi ^{2}\omega _{2}^{2} \right) \omega _{2}
\nonumber
\\
0 & = & 
{\ddot {\psi }}-{\dot {\psi }}+\frac {2{\dot {\omega }}_{2}}{
\omega _{2}}{\dot {\psi }}+\frac {3}{2} \omega _{2}^{2}
\psi \left( 1-\psi ^{2} \right).
\eea
Without loss of generality, we may assume that we are close to the
critical point where $\psi =1$ and $\omega _{2}={\sqrt {2}}$.
Using the substitution
\be
\psi = 1+\epsilon _{1},
\qquad
\omega _{2}={\sqrt {2}}(1+\epsilon _{2}),
\ee
the linearized equations close to this critical point are:
\bea
0 & = & 
{\ddot {\epsilon }}_{1}-{\dot {\epsilon }}_{1} 
-6\epsilon _{1}
\nonumber
\\
0 & = & 
{\ddot {\epsilon }}_{2}-{\dot {\epsilon }}_{2} 
+2\epsilon _{1}-2\epsilon _{2}.
\eea
The analysis of section \ref{flat} has already shown that we have
a saddle point here; this notation is convenient merely because
the perturbation in $\psi $ decouples from that of $\omega _{2}$
in the linear approximation.
From the form of the equation (\ref{psieqn}), it is clear that
$|\psi |$ will be monotonic increasing for $|\psi |>1$.
In other words, for $\epsilon _{1}>0$ initially, $\epsilon _{1}>0$
always and $\omega _{1}$ will have no additional zeros for $r>r_{1}$.
Hence we need only consider the case where $\epsilon _{1}<0$
initially.
Unfortunately the non-linear perturbation equations cannot be
integrated analytically and so we present a numerical argument.
For initial $\epsilon _{1},\epsilon _{2}$ sufficiently small,
the values of $\epsilon _{1}$ and $\epsilon _{2}$ will remain close
to those on the unstable manifold for all $\tau <\tau _{1}$ for
some $\tau _{1}$, where $\tau _{1}$ can be taken to be as large
as we like by making the initial perturbations sufficiently small.
Numerical integration of the non-linear equations with the 
initial point on the unstable manifold and $\epsilon _{1}<0$,
shows that $\epsilon _{2}$ increases monotonically and 
$\epsilon _{1}$ decreases monotonically with just one zero, and 
cuts through $\epsilon _{1}=-2$ (corresponding to $\psi =-1$),
from whence $\psi $ must be monotonically decreasing
(see figures 4 and 5).
Hence in this case $\omega _{1}$ has ${\bar {n}}_{1}+1$ zeros.
\hfill
$\square $ 

\bigskip
Lemma \ref{tightlem} allows us to prove that every combination
of integers $(n_{1},n_{2})$ must correspond to a black hole
solution.
In the proof of theorem \ref{exist}, we begin with regular
solutions with node structure $({\bar {n}}_{2},{\bar {n}}_{2})$
and then from lemma \ref{tightlem} there are either regular 
or singular solutions with node structure
$({\bar {n}}_{2}+1,{\bar {n}}_{2})$.
Using lemma \ref{singlem}, there are then regular solutions
having this node structure and the proof of theorem \ref{exist}
follows as before. 

\bigskip
So far we have proved the existence of infinitely many 
${\mathfrak {su}}(3)$ black holes only for $r_{h}>2$.
The remainder of this section will be spent proving the result
for $r_{h}\le 2$.
In this case, the points at which the map $f$ is not continuous
(which must still exist by the argument used in proving proposition
\ref{firstprop}) do not necessarily correspond to regular black hole
solutions: they could also be oscillating solutions.
A couple of lemmas concerning oscillating solutions are required
before the proofs of proposition \ref{firstprop} and theorem
\ref{exist} can be extended.

\begin{lemma}
\label{osclem}
If $({\bar {\omega }}_{1,h},{\bar {\omega }}_{2,h})$
leads to an oscillating solution with 
$\omega _{1}(\tau )\rightarrow 0$ and 
$\omega _{2}(\tau )\rightarrow 1$ as 
$\tau \rightarrow \infty $, and
$\omega _{2}$ has ${\bar {n}}_{2}$ zeros, then all 
$(\omega _{1,h},\omega _{2,h})$ sufficiently close to
$({\bar {\omega }}_{1,h},{\bar {\omega }}_{2,h})$
lead to one of the following types of solution:
\begin{enumerate}
\item
an oscillating solution in which $\omega _{1}(\tau )\rightarrow 0$ and 
$\omega _{2}(\tau )\rightarrow 1$ as 
$\tau \rightarrow \infty $, with $\omega _{2}$ having at least
${\bar {n}}_{2}$ zeros;
\item
given $n_{0}$, either a regular or a singular solution with
$n_{1}\ge n_{0}$ and $n_{2}\ge {\bar {n}}_{2}$.
\item
an $S_{\infty }$ solution.
\end{enumerate}
\end{lemma}

\proof
Since in the original solution $\omega _{1}$ has infinitely many
zeros and the solutions are analytic in $\tau $ and the starting
values, there is some $\tau _{1}$ such that 
all solutions with $(\omega _{1,h},\omega _{2,h})$ 
sufficiently close to $({\bar {\omega }}_{1,h},{\bar {\omega }}_{2,h})$
must have at least $n_{0}$ zeros of $\omega _{1}$ and 
${\bar {n}}_{2}$ zeros of $\omega _{2}$ for $\tau <\tau _{1}$,
and all variables will be close to their asymptotic values.
\hfill
$\square $

\begin{lemma}
There exists $({\bar {\omega }}_{1,h},{\bar {\omega }}_{2,h})\in D$
corresponding to an oscillating solution.
\end{lemma}

\proof
From \cite{bfm}, we know that there are points on the line
$\omega _{1,h}=\omega _{2,h}$ corresponding to $S_{\infty }$
solutions.
By continuity, there is a neighbourhood $D_{0}$ of the line
$\omega _{1,h}=\omega _{2,h}$ corresponding to $S_{\infty }$
solutions.
However, lemma \ref{reglem} still applies in this case, 
so that the regular solutions on the line 
$\omega _{1,h}=\omega _{2,h}$ will have singular solutions in $D$
sufficiently close to them for which $r(\tau _{0})>2$.
Therefore, by lemma \ref{osclem}, there must be points in $D$
which correspond to oscillating solutions.
\hfill
$\square $

\bigskip
The proofs of proposition \ref{firstprop} and theorem \ref{exist}
are now exactly the same as for $r_{h}>2$. 
Since oscillating solutions
have an infinite number of zeros for at least one of the gauge field
functions, and using lemma \ref{osclem},
regular solutions are still needed to separate 
singular solutions having different node structures.

\section{${\mathfrak {su}}(N)$ black holes}
\label{sun}
The detailed discussion of the previous section now enables us to
proceed quite swiftly to the analogues of proposition \ref{firstprop}
and theorem \ref{exist} for general $N$. 
The method of proof will be by induction, which was the basic idea used 
in the previous section, where we proved existence for 
${\mathfrak {su}}(3)$ exploiting known results about 
${\mathfrak {su}}(2)$.

\bigskip
The first step is to illustrate how solutions which are from 
${\mathfrak {su}}(n)$, where $n<N$ may be embedded in the 
${\mathfrak {su}}(N)$ framework.
Firstly, exactly as in the $N=3$ case, we may embed neutral
${\mathfrak {su}}(2)$ solutions \cite{kunz1}.
We set $\omega _{j}(r)={\sqrt {j(N-j)}}\omega (r)$ and introduce
a scaled variable $R=\lambda _{N}r$, where
\be
\lambda _{N}=\left( \frac {1}{6} N(N+1)(N-1) \right) ^{\frac {1}{2}}.
\ee
Then the field equations are exactly the same as the
${\mathfrak {su}}(2)$ ones, and the existence of solutions follows 
directly \cite{kunz1}.
Secondly, charged, effectively ${\mathfrak {su}}(N-1)$ solutions 
can be generated by setting $\omega _{1}\equiv 0$ (or
$\omega _{N-1}\equiv 0$) \cite{klei1}.
Although the equation for $m'$ in this situation is not the same as
for neutral ${\mathfrak {su}}(N-1)$ black holes, due to the charge, 
the regular equations discussed in section \ref{ansatz} are unchanged,
so the proof of existence of neutral ${\mathfrak {su}}(N-1)$
black holes extends naturally to this case.

\bigskip
For $N>3$, there are additional embeddings which arise from setting
at least one of $\omega _{2}, \ldots , \omega _{N-2}\equiv 0$,
so that the gauge field equations decouple into two or more
sets of coupled components, the sets being coupled to each other 
only through the metric.  Solutions of this form have been found
numerically in \cite{klei1}.
We illustrate how the existence of black holes of this type may
be proved by considering the simplest case, which arises when
$N=4$.
In this case, we have three non-zero gauge field functions, 
$\omega _{1},\omega _{2}, \omega _{3}$.
If we set $\omega _{2}\equiv 0$, then the field equations take
the form
\bea
r^{2}\mu \omega _{1}'' & = & 
-\left( 2m-2r^{3}p_{\theta } \right) \omega _{1}'
-\left( 1-\omega _{1}^{2} \right) \omega _{1}  
\label{omone}
\\
r^{2}\mu \omega _{3}'' & = & 
-\left( 2m-2r^{3}p_{\theta } \right) \omega _{3}'
-\left( 1-\omega _{3}^{2} \right) \omega _{3} 
\label{omtwo}
\\
m' & = & \left( \mu G +r^{2} p_{\theta }\right)
\\
\frac {S'}{S} & = & \frac {2G}{r}
\eea
where
\be
G=\omega _{1}^{'2} + \omega _{3}^{'2},
\qquad
p_{\theta } = \frac {1}{2r^{4}} \left( 
\left[ \omega _{1}^{2}-1 \right] ^{2}+ 
\left[ \omega _{3}^{2}-1 \right] ^{2}+ 8 \right) .
\ee
These equations look very much like two uncoupled ${\mathfrak {su}}(2)$
degrees of freedom, however, the two $\omega $'s are (albeit weakly)
coupled.
The fact that we have two gauge field functions here means that we
can use the methods of section \ref{su3} to prove the existence of
regular black hole solutions.
In fact, all the results of that section carry directly over to this
situation on replacing $\omega _{2}$ there by $\omega _{3}$ here.
There is one exception. 
The equations (\ref{omone},\ref{omtwo}) are slightly different from
those in section \ref{su3} and this enables us to strengthen
lemma \ref{reglem}.

\begin{lemma}
If $({\bar {\omega }}_{1,h},{\bar {\omega }}_{3,h})$
leads to a regular black hole solution with $\omega _{1}$ having
${\bar {n}}_{1}$ nodes and $\omega _{3}$ having ${\bar {n}}_{3}$
nodes, then all $(\omega _{1,h},\omega _{3,h})$ sufficiently 
close to
$({\bar {\omega }}_{1,h},{\bar {\omega }}_{3,h})$
lead to regular or singular solutions with $\omega _{1}$ having
either ${\bar {n}}_{1}$ or ${\bar {n}}_{1}+1$ zeros and 
$\omega _{3}$ having either ${\bar {n}}_{3}$ or ${\bar {n}}_{3}+1$
zeros.
\end{lemma}

\proof
Since the solutions are continuous in the starting parameters and $r$,
for $(\omega _{1,h},\omega _{3,h})$ sufficiently close to
$({\bar {\omega }}_{1,h},{\bar {\omega }}_{3,h})$,
there is some $r_{1}$ such that $\omega _{1}$ will have ${\bar {n}}_{1}$
zeros and $\omega _{3}$ will have ${\bar {n}}_{3}$ zeros for
$r<r_{1}$. 
There will also be an $r_{2}>r_{1}$ such that all field variables
are close to their asymptotic values for $r\in [r_{1},r_{2}]$.
The flat space field equations (see section \ref{flat}) decouple
completely since $\omega _{2}\equiv 0$, and so the phase plane
analysis of \cite{bfm} is valid in this case.
The phase plane analysis enables us to be more precise about
the number of zeros of the perturbed gauge field functions, 
unlike the more general case where the phase space had more than
two coupled dimensions.
Therefore the conclusions of \cite[proposition 22]{bfm} can 
be applied directly here and each gauge field function
can have at most one zero for $r>r_{2}$.
\hfill
$\square $

\bigskip
With this more powerful lemma, the analysis of section \ref{su3}
reaches the conclusion that there are regular black hole
solutions of the required form for each $r_{h}$ and node structure
$(n_{1},n_{3})$.

\bigskip
For general $N$, similar embeddings of the form of two or more
${\mathfrak {su}}(n)$ ($n<N$) type solutions separated by at least
one $\omega _{j}\equiv 0$ are possible, and the existence proof
follows the lines outlined above for ${\mathfrak {su}}(4)$, 
using the existence of solutions for the various 
${\mathfrak {su}}(n)$.
The allowed node structures will be exactly the same as those for the
constituent ${\mathfrak {su}}(n)$ components.
We refer the reader to \cite{klei1} for further details of the
construction of solutions of this form.

\bigskip
The remainder of this section will be spent outlining briefly the
proof of the following theorem.
By `types' in the statement of the theorem we mean the node structures
of the gauge field functions.

\begin{theorem}
\label{nexist}
There exist infinitely many types of regular black hole
solutions of the ${\mathfrak {su}}(N)$ Einstein-Yang-Mills
equations including genuine ${\mathfrak {su}}(N)$ solutions and 
embedded solutions.
\end{theorem}

All that remains to prove this theorem is to show the existence
of genuine ${\mathfrak {su}}(N)$ solutions in which the node
structures of the gauge field functions are not all the same.
We assume for an inductive hypothesis that the theorem
holds for all $n<N$.
The situation is slightly complicated by the fact that our 
parameter space consisting of the $\omega _{j,h}$'s is now
$N-1$-dimensional.
We consider the $\omega _{1,h}=0$ hyperplane, on which we know
the solution space by the inductive hypothesis.
The map $f$ (see section \ref{su3}) now takes ${\mathbb {R}}^{N-1}$
to ${\mathbb {B}}^{N-1}\times [0,1]$.
Again, we need only consider the action of $f$ on those parts of
${\mathbb {R}}^{N-1}$ in which no $\omega _{j,h}$ vanishes.
We restrict attention to one of the disjoint spaces so generated 
without loss of generality.
Lemma \ref{fcont} and the proof of the ${\mathfrak {su}}(N)$
equivalent of proposition \ref{firstprop} now carry over to 
prove the existence of genuine ${\mathfrak {su}}(N)$ black holes.
Lemmas \ref{weaklem}, \ref{singlem} and \ref{existlem}
 now hold  for a point 
$(0,{\bar {\omega }}_{2,h}, \ldots ,{\bar {\omega }}_{N-1,h})$
corresponding to a charged solution, and lemmas \ref{reglem}
and \ref{osclem} carry straight over to a point
$({\bar {\omega }}_{1,h},{\bar {\omega }}_{2,h}, \ldots ,
{\bar {\omega }}_{N-1,h})$ which leads to a regular black hole
solution or oscillating solution respectively.
These are the ingredients necessary for the simple argument which
leads to the proof of theorem \ref{exist} and hence theorem
\ref{nexist} for both $r_{h}^{2}>\frac {1}{6}N(N+1)(N-1)$ (in which
case no oscillating solutions exist), and 
$r_{h}^{2}\le \frac {1}{6} N(N+1)(N-1)$.
Note that, as in the ${\mathfrak {su}}(3)$ case, we are not able to
prove analytically that every sequence of integers 
$(n_{1},n_{2},\ldots ,n_{N-1})$ corresponds to a black hole solution
whose gauge functions have this node structure, although it might
reasonably be expected that this is indeed the case.


\section{Results and conclusions}
In this paper we have proved the existence of a vast number of
hairy black holes in ${\mathfrak {su}}(N)$ Einstein-Yang-Mills
theories.
The solutions are described by N-1 parameters,
corresponding to the number of nodes of the gauge field
functions.
The result of \cite{brod} tells us that each of these solutions
will have a topological instability, similar to the flat-space
``sphaleron'', as was the case for ${\mathfrak {su}}(2)$
black holes.
This instability does
not necessarily diminish the physical importance of these objects,
as they may have an important role in processes such as cosmological
particle creation \cite{gibbons}.

\bigskip
We now briefly mention another important reason for studying
${\mathfrak {su}}(N)$ black holes, namely the behaviour
of the solutions as $N\rightarrow \infty $.
Black holes in ${\mathfrak {su}}(\infty )$
Einstein-Yang-Mills theory would possess an infinite amount
of hair, requiring an infinite number of parameters to
describe their geometry.
This might be analogous to the ``W-hair'' found in 
non-critical string theory \cite{ellis}, and have 
drastic consequences for the Hawking radiation,
information loss and quantum decoherence processes
associated with black holes.
The existence of infinite amounts of hair might also 
render such objects stable.
It is already known that the ${\mathfrak {su}}(\infty )$
Lie algebra is simply that of the diffeomorphisms of the sphere
\cite{floratos}.
This means that the limit as $N\rightarrow \infty $ cannot
be taken smoothly, in particular the results of the present paper
are only valid when $N$ is finite.
We hope to return to this matter in a subsequent publication.

\section*{Acknowledgments}
The work of N.E.M. is supported by a P.P.A.R.C. advanced fellowship
and that of E.W. is supported by a fellowship at Oriel College, Oxford.

\newpage

\begin{figure}
\unitlength 1.00mm
\linethickness{0.4pt}
\begin{picture}(135.00,130.00)
\put(130.00,20.00){\line(0,1){0.00}}
\put(14.66,120.00){\makebox(0,0)[cc]{$\omega _{2,h}$}}
\put(120.33,15.00){\makebox(0,0)[cc]{$\omega _{1,h}$}}
\put(135.00,120.00){\makebox(0,0)[cc]{$\omega _{1,h}=\omega _{2,h}$}}
\put(20.00,20.00){\line(1,0){110.00}}
\put(130.00,20.00){\line(-1,0){110.00}}
\put(20.00,20.00){\line(1,1){110.00}}
\put(20.00,20.00){\line(0,1){110.00}}
\put(75.00,75.00){\makebox(0,0)[cc]{$\times $}}
\put(75.00,20.00){\makebox(0,0)[cc]{$\times $}}
\put(20.00,75.00){\makebox(0,0)[cc]{$\times $}}
\put(50.00,50.00){\makebox(0,0)[cc]{$\times $}}
\put(20.00,50.00){\makebox(0,0)[cc]{$\times $}}
\put(50.00,20.00){\makebox(0,0)[cc]{$\times $}}
\put(35.00,35.00){\makebox(0,0)[cc]{$\times $}}
\put(20.00,35.00){\makebox(0,0)[cc]{$\times $}}
\put(35.00,20.00){\makebox(0,0)[cc]{$\times $}}
\bezier{364}(60.00,110.00)(65.00,65.00)(110.00,60.00)
\bezier{848}(45.00,120.00)(17.67,17.33)(120.00,45.00)
\bezier{100}(30.00,45.00)(32.67,32.33)(45.00,30.00)
\end{picture}
\caption{The $(\omega _{1,h},\omega _{2,h})$ plane, with a schematic 
representation of the lines representing regular black hole solutions,
which are also denoted by crosses.}
\end{figure}
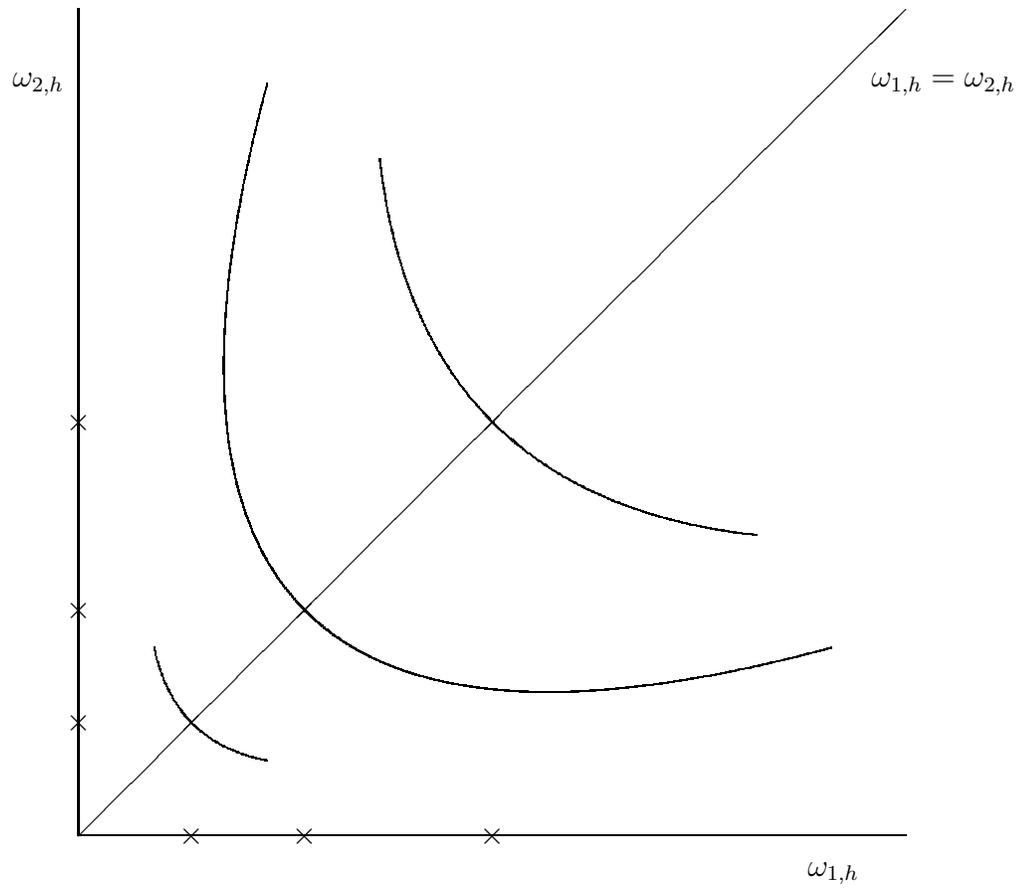

\begin{figure}
\unitlength 1.00mm
\linethickness{0.4pt}
\begin{picture}(100.00,100.00)
\put(25.33,99.33){\makebox(0,0)[rc]{$R_{m}=1$}}
\put(25.33,30.00){\makebox(0,0)[rc]{$R_{m}=0$}}
\put(30.00,25.00){\makebox(0,0)[lc]{$N_{1}=0$}}
\put(65.00,25.00){\makebox(0,0)[lc]{$N_{1}=0.5$}}
\put(100.00,25.00){\makebox(0,0)[lc]{$N_{1}=1$}}
\put(100.00,30.00){\line(-1,0){73.00}}
\put(100.00,100.00){\line(-1,0){73.00}}
\put(100.00,100.00){\line(0,-1){73.00}}
\put(30.00,100.00){\line(0,-1){73.00}}
\put(65.00,100.00){\line(0,-1){73.00}}
\put(76.00,100.00){\line(0,-1){70.00}}
\put(82.00,100.00){\line(0,-1){70.00}}
\put(86.00,100.00){\line(0,-1){70.00}}
\put(88.00,100.00){\line(0,-1){70.00}}
\put(90.00,100.00){\line(0,-1){70.00}}
\end{picture}
\caption{Schematic representation of the set ${\mathbb {B}}\times
[0,1]$,
corresponding to the space of parameters of the 
${\mathfrak {su}}(3)$ solutions in section VII.
The horizontal axis represents $N_{1}$, the number of nodes of the
gauge function $\omega _{1}$, and the vertical axis $R_{m}$, 
which characterizes the maximum value of the radial co-ordinate $r$
for each solution.}
\end{figure}
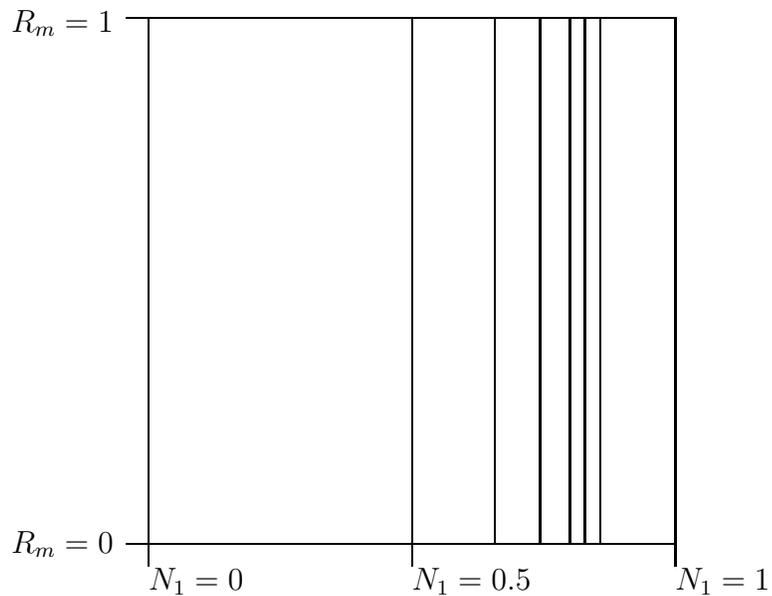

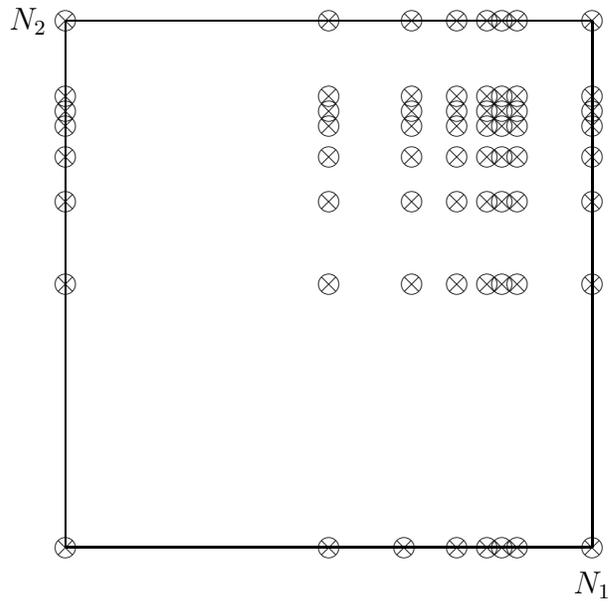
\begin{figure}
\unitlength 1.00mm
\linethickness{0.4pt}
\begin{picture}(100.00,100.00)
\put(25.00,100.00){\makebox(0,0)[cc]{$N_{2}$}}
\put(100.00,25.00){\makebox(0,0)[cc]{$N_{1}$}}
\put(30.00,30.00){\makebox(0,0)[cc]{$\otimes $}}
\put(65.00,30.00){\makebox(0,0)[cc]{$\otimes $}}
\put(75.00,30.00){\makebox(0,0)[cc]{$\otimes $}}
\put(82.00,30.00){\makebox(0,0)[cc]{$\otimes $}}
\put(86.00,30.00){\makebox(0,0)[cc]{$\otimes $}}
\put(88.00,30.00){\makebox(0,0)[cc]{$\otimes $}}
\put(90.00,30.00){\makebox(0,0)[cc]{$\otimes $}}
\put(100.00,30.00){\makebox(0,0)[cc]{$\otimes $}}
\put(30.00,65.00){\makebox(0,0)[cc]{$\otimes $}}
\put(65.00,65.00){\makebox(0,0)[cc]{$\otimes $}}
\put(76.00,65.00){\makebox(0,0)[cc]{$\otimes $}}
\put(82.00,65.00){\makebox(0,0)[cc]{$\otimes $}}
\put(86.00,65.00){\makebox(0,0)[cc]{$\otimes $}}
\put(88.00,65.00){\makebox(0,0)[cc]{$\otimes $}}
\put(90.00,65.00){\makebox(0,0)[cc]{$\otimes $}}
\put(100.00,65.00){\makebox(0,0)[cc]{$\otimes $}}
\put(30.00,76.00){\makebox(0,0)[cc]{$\otimes $}}
\put(65.00,76.00){\makebox(0,0)[cc]{$\otimes $}}
\put(76.00,76.00){\makebox(0,0)[cc]{$\otimes $}}
\put(82.00,76.00){\makebox(0,0)[cc]{$\otimes $}}
\put(86.00,76.00){\makebox(0,0)[cc]{$\otimes $}}
\put(88.00,76.00){\makebox(0,0)[cc]{$\otimes $}}
\put(90.00,76.00){\makebox(0,0)[cc]{$\otimes $}}
\put(100.00,76.00){\makebox(0,0)[cc]{$\otimes $}}
\put(30.00,82.00){\makebox(0,0)[cc]{$\otimes $}}
\put(65.00,82.00){\makebox(0,0)[cc]{$\otimes $}}
\put(76.00,82.00){\makebox(0,0)[cc]{$\otimes $}}
\put(82.00,82.00){\makebox(0,0)[cc]{$\otimes $}}
\put(86.00,82.00){\makebox(0,0)[cc]{$\otimes $}}
\put(88.00,82.00){\makebox(0,0)[cc]{$\otimes $}}
\put(90.00,82.00){\makebox(0,0)[cc]{$\otimes $}}
\put(100.00,82.00){\makebox(0,0)[cc]{$\otimes $}}
\put(30.00,86.00){\makebox(0,0)[cc]{$\otimes $}}
\put(65.00,86.00){\makebox(0,0)[cc]{$\otimes $}}
\put(76.00,86.00){\makebox(0,0)[cc]{$\otimes $}}
\put(82.00,86.00){\makebox(0,0)[cc]{$\otimes $}}
\put(86.00,86.00){\makebox(0,0)[cc]{$\otimes $}}
\put(88.00,86.00){\makebox(0,0)[cc]{$\otimes $}}
\put(90.00,86.00){\makebox(0,0)[cc]{$\otimes $}}
\put(100.00,86.00){\makebox(0,0)[cc]{$\otimes $}}
\put(30.00,88.00){\makebox(0,0)[cc]{$\otimes $}}
\put(65.00,88.00){\makebox(0,0)[cc]{$\otimes $}}
\put(76.00,88.00){\makebox(0,0)[cc]{$\otimes $}}
\put(82.00,88.00){\makebox(0,0)[cc]{$\otimes $}}
\put(86.00,88.00){\makebox(0,0)[cc]{$\otimes $}}
\put(88.00,88.00){\makebox(0,0)[cc]{$\otimes $}}
\put(90.00,88.00){\makebox(0,0)[cc]{$\otimes $}}
\put(100.00,88.00){\makebox(0,0)[cc]{$\otimes $}}
\put(30.00,90.00){\makebox(0,0)[cc]{$\otimes $}}
\put(65.00,90.00){\makebox(0,0)[cc]{$\otimes $}}
\put(76.00,90.00){\makebox(0,0)[cc]{$\otimes $}}
\put(82.00,90.00){\makebox(0,0)[cc]{$\otimes $}}
\put(86.00,90.00){\makebox(0,0)[cc]{$\otimes $}}
\put(88.00,90.00){\makebox(0,0)[cc]{$\otimes $}}
\put(90.00,90.00){\makebox(0,0)[cc]{$\otimes $}}
\put(100.00,90.00){\makebox(0,0)[cc]{$\otimes $}}
\put(30.00,100.00){\makebox(0,0)[cc]{$\otimes $}}
\put(65.00,100.00){\makebox(0,0)[cc]{$\otimes $}}
\put(76.00,100.00){\makebox(0,0)[cc]{$\otimes $}}
\put(82.00,100.00){\makebox(0,0)[cc]{$\otimes $}}
\put(86.00,100.00){\makebox(0,0)[cc]{$\otimes $}}
\put(88.00,100.00){\makebox(0,0)[cc]{$\otimes $}}
\put(90.00,100.00){\makebox(0,0)[cc]{$\otimes $}}
\put(100.00,100.00){\makebox(0,0)[cc]{$\otimes $}}
\put(30.00,100.00){\line(1,0){70.00}}
\put(100.00,100.00){\line(0,-1){70.00}}
\put(100.00,30.00){\line(-1,0){70.00}}
\put(30.00,30.00){\line(0,1){70.00}}
\end{picture}
\caption{The projection of the parameter
space of solutions on to the set ${\mathbb {B}}\times {\mathbb {B}}$.
The axes are $N_{1}$ and $N_{2}$, the number of nodes of the 
respective gauge functions $\omega _{1}$ and $\omega _{2}$.
The crosses indicate schematically the points in this set.}
\end{figure}

\begin{figure}
\begin{center}
\includegraphics[width=12cm]{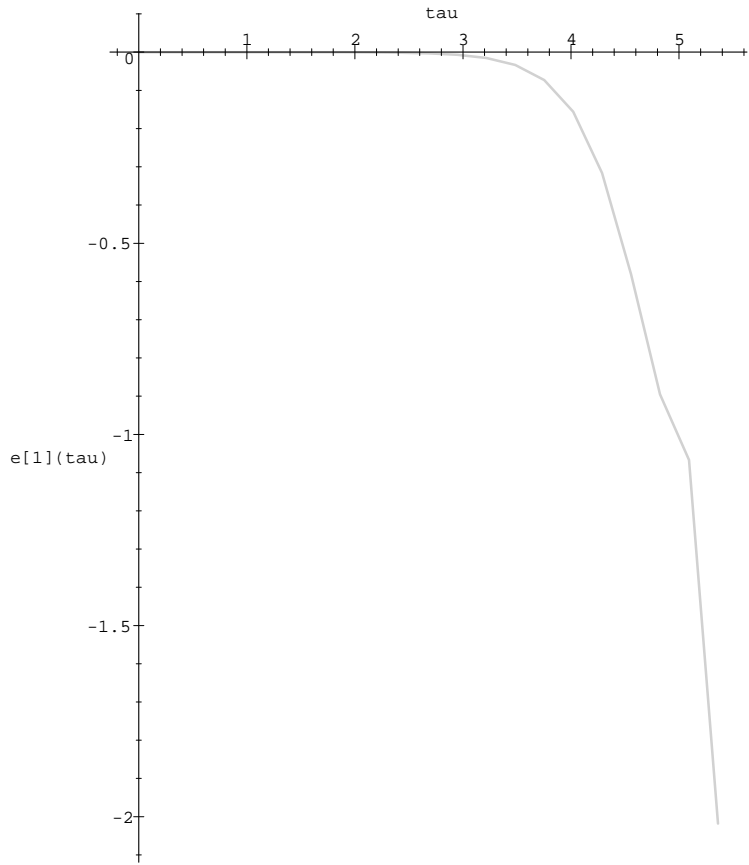}
\end{center}
\caption{The ${\mathfrak {su}}(3)$
perturbation $\epsilon _{1}(\tau )$ along the
unstable manifold. The value 
$\tau =0$ corresponds to the critical point.  Note that 
$\psi =1+\epsilon _{1}$ monotonically decreases through $0$ and $-1$,
from which point it will continue to monotonically decrease.}
\end{figure}

\begin{figure}
\begin{center}
\includegraphics[width=12cm]{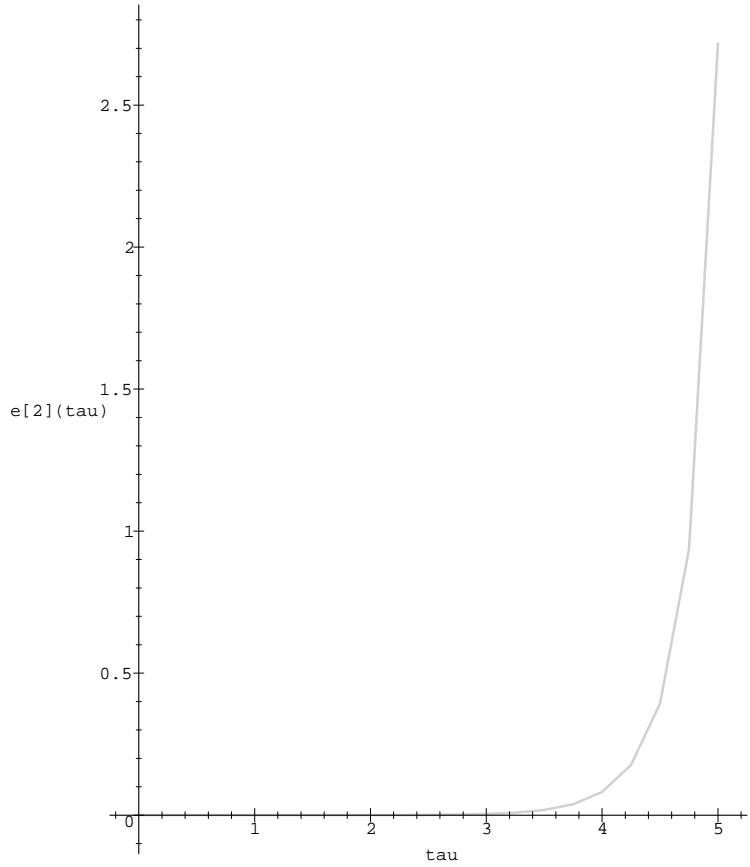}
\end{center}
\caption{The ${\mathfrak {su}}(3)$ perturbation
$\epsilon _{2}$ along the unstable manifold.
Note that this is a monotonically increasing function, which is in 
accordance with the gauge function $\omega _{2}$ having no further 
zeros.}
\end{figure}

\end{document}